\documentclass[twocolumn]{revtex4}
\usepackage{epsfig,bm,dcolumn}
\usepackage{graphicx}
\usepackage{epsfig}
\usepackage{latexsym}
\usepackage{bm}
\usepackage{amsmath}
\usepackage{lipsum}
\usepackage{graphicx}
\usepackage{dcolumn}
\usepackage{float}
\usepackage{bm}
\usepackage{gensymb}
\usepackage{amssymb}

\begin{document}

\preprint{APS/MI}

\title{Interplay of mass imbalance and frustration in correlated band insulators}

\author{Anwesha Chattopadhyay}
 \affiliation{Condensed Matter Physics Division, Saha Institute of Nuclear Physics, HBNI, 1/AF Bidhannagar, Kolkata 700 064, India}

\date{\today}

\begin{abstract}
We report the emergence or broadening of exotic magnetic metallic phases upon explicit breaking of $SU(2)$ symmetry by introduction of mass imbalance in a variant of Hubbard model, known as the ionic Hubbard model in the presence of frustration at half-filling on a square lattice. The ionic Hubbard model has in addition to hopping($\sim t$) and onsite coulomb repulsion ($\sim U$), a staggered ionic potential ($\sim \Delta$) which breaks translational symmetry of the underlying lattice. In the low to intermediate ranges of $U$ and $\Delta$, we use unrestricted Hartree-Fock theory to construct the phase diagram in the $U-\Delta$ plane for a fixed mass imbalance($\sim\eta$). Where as in the limit where both $U,\Delta$ are comparable and much larger than the first $(\sim t_{\sigma})$ and second $(\sim t_{\sigma}')$ neighbor hopping amplitudes , we employ the technique of generalized Gutzwiller approximation and subsequently use renormalized mean field theory to construct the phase diagram for varying values of hopping asymmetry. In both cases, starting from a correlated band insulator with weak antiferromagnetic spin density wave order, if we tune $U/\Delta$, we observe the opening of novel magnetically ordered metallic phases such as spin imbalanced ferromagnetic metal, ferrimagnetic metal and antiferromagnetic half metal. We also study singlet superconductivity in the d-wave and extended s-wave channels  in the strong coupling limit of this model in the presence of mass imbalance.
\end{abstract}

\maketitle


\section{Introduction}

Many condensed matter systems show rich phenomenology like high $T_{c}$ superconductivity, Mott insulating properties, spin liquid behaviour etc. An idealistic model which accounts for these phenomena in atleast a qualitative way is the Hubbard model (HM)~\cite{Hubbard}. In principle, the HM captures many non-trivial physical phenomena  which are also found in realistic complicated systems which belong to the same ``Universality class". An interesting variant of the Hubbard model is the ionic Hubbard model (IHM) which has an onsite Coulomb interaction term ($\sim U$) and a staggered ionic potential term ($\sim \Delta$) applied to itinerant electrons. The IHM has been studied extensively in the past using dynamical mean field theory (DMFT)~\cite{Jabben,Garg1,Craco,Byczuk,Garg2,Wang1,Kim,Bag1,Bag2}, determinantal quantum Monte Carlo (DQMC)~\cite{Paris,Bouadim}, cluster DMFT~\cite{Kancharla}, density matrix renormalization group (DMRG)~\cite{Manmana}, coherent potential approximation~\cite{Hoang} and renormalized mean field theory (RMFT)~\cite{Anwesha1,Anwesha2,Anwesha3} .While $U$ prefers single occupancies on sites, $\Delta$ prefers a staggered charge density ordering. The competition between these insulating tendencies can give rise to interesting metallic, half-metallic, bond-ordered or superconducting phases ~\cite{Garg1,Craco,Paris,Bouadim,Hoang,Garg2,Kancharla,Anwesha3}.

An interesting question to ask is what will happen if we introduce hopping asymmetry in the IHM. Introducing mass imbalanced ultracold fermionic species like ${}^{6}Li,{}^{40}K$ on an optical lattice and tuning inter-species Coulomb interaction via magnetic Feshbach resonance can provide us an ideal platform for investigation in this direction. Moreover, the staggered ionic potential can be created by interference of counter propagating laser beams. Indeed, IHM was experimentally realized on an optical honeycomb lattice, but using single species ultracold fermionic atoms~\cite{IHMexpt}. A state dependent optical lattice was realized for ${}^{40}K$ fermionic atoms by using magnetic field gradient modulated in time which tuned the relative amplitude and sign of the hopping depending on the internal spin state~\cite{Jotzu}. This allows for the investigation of effective mass imbalanced physics by using only single species fermionic atoms. Some experimental studies have also been reported where mass imbalanced fermi-fermi atomic mixtures have been successfully realized~\cite{Taglieber,Wille, Spiegelhalder,Trenkwalder,Kohstall,Jag}. Also, mass imbalanced fermionic species have been studied theoretically in different contexts including superfluidity, Mott transition etc~\cite{Dao1,Wang2,Gubbels,Baarsma,Dao2,Takemori,Sotnikov,Hanai1,Lan,Hanai2,Gukelberger,Nguyen,Hu,Liu1,Wang3,Philipp,Le,Muller}. Further, effective mass imbalance can also be achieved in solid-state systems  having bands of varying bandwidths which  can show orbital selective Mott transition~\cite{Koga,Liebsch,Medici,Ferrero,Aritra}. Moreover, mass imbalance is tunable in optical lattices by changing the lattice depth of the optical trap. Infact, the lattice depth and hence the hopping amplitude can be different for same species components in different hyperfine levels without mass imbalnce~\cite{Liu2}.  

We now look into the aspect of frustration in the context of IHM. Often, strong quantum fluctuations make a critical phase unstable or metastable. These phases can be stabilized by introducing frustation in the system~\cite{Frust}. Frustation can be geometric in nature or can be due to the introduction of next nearest neighbor hopping (for e.g., on a square lattice). Two recent studies~\cite{Bag2,Anwesha3} exploit this property of frustration in IHM for stabilizing intermediate phases. In~\cite{Bag2}, IHM is solved with finite second neighbor hopping at half-filling on a square lattice using DMFT with CTQMC impurity solver and Hartree Fock mean field theory where the authors find correlation induced paramagnetic (PM) metallic, ferrimagentic metallic and antiferromagnetic (AF) half metallic phases starting from a band insulator. The correlation induced metallic phase was observed earlier in a DMFT study~\cite{Garg1} (with first neighbor hopping) in the PM sector at half-filling. However, when AF order was allowed, instead of an intervening metallic phase, a direct transition from band insulator to AF insulator was observed~\cite{Byczuk,Kancharla}. The other work~\cite{Anwesha3}, co-authored by us, shows  unconventional superconductivity at half-filling in the IHM with first ($\sim t$) and second ($\sim t'$) neighbor hoppings on a square lattice in the regime where $U\sim \Delta \gg t,t'$  using a generalization of projected wave functions method.  

In this paper, we study the interplay between mass imbalance and frustration in the IHM at half-filling on a square lattice.
Specifically, we study the model for low to intermediate values of $U,\Delta$ using unrestricted Hartree-Fock theory and in the limit $U\sim \Delta \gg t_{\sigma},t_{\sigma}'$ we use a generalization of Gutzwiller approximation~\cite{Gutzwiller1,Gutzwiller2,Gutzwiller3} following which renormalized mean field theory is used to develop the phase diagram. The IHM was studied with mass imbalance in the presence of nearest neighbor hopping featuring a weakly first order transition from a predominantly density modulated phase to a predominantly spin alternating phase~\cite{Sekania}. For all $U\geqslant 0$, period 2 charge modulation and alternating spin density was found to coexist. However with second neighbor hopping, we observe exotic metallic phases which are symmetry broken and have potential applications in the field of spintronics~\cite{Wolf}.  The ferrimagnetic metallic phase, earlier observed in the mass balanced frustrated IHM in a relatively narrow regime of the phase space ~\cite{Bag2} is broadened significantly in the presence of mass imbalance. Further, we also find AF half metallic phases in both the limits studied. Specially, in the strong coupling limit we find both type of spin polarized AF half-metals which can be switched by tuning $U/\Delta$. However, the most interesting and novel phase that we find is that of a {\it spin imbalanced ferromagnetic} metallic phase where the spins on alternate sites are oriented in the same direction but differ in magnitude between sublattices, similar to ferrimagnetic phase where the spins on different sublattices are oriented opposite to each other but with different magnitude. According to Stoner criteria~\cite{Fazekas,Tasaki}, ferromagnetism  can arise if $UD(\epsilon_{F})>1$ where $U$ is the Coulomb interaction and $D(\epsilon_{F})$ is the density of states at the fermi level for non-interacting electrons. Further, according to Nagaoka~\cite{Fazekas,Tasaki}, in the $U \rightarrow \infty$ limit, the HM admits a ferromagnetic ground state in the presence of exactly one hole in the system. Also, the system can show flat band ferromagnetism~\cite{Fazekas,Tasaki} if a dispersionless band exists at the bottom of the spectra. Both Nagaoka and flat band ferromagnetism occur when $U/\Delta E \rightarrow \infty$, $\Delta E$ being the non-interacting bandwidth, either because $U/t \rightarrow \infty$ or relevant band is flat. The existence of ferromagnetism at finite values of $U/\Delta E$ was first addressed by Tasaki~\cite{Fazekas,Tasaki} who showed ferromagnetism can be stabilized at non-singular values by ring exchange mechanism where triangular plaquettes connecting sites through non zero hopping is essential. This indicates the necessity of introducing frustration in the system for stabilizing ferromagnetic phases by killing AF order. Also since mass imbalance effectively reduces the bandwidth of the heavier species
creating almost flat bands, the divergence of density of states at the fermi level is expected to create metallic ferromagnetism in the system. Experimentally, itinerant ferromagnetism has been observed in fermi gases of ultracold atoms which are in two different pseudo spin levels~\cite{Jo}. Subsequently, theoretically a mass imbalanced fermi gas with repulsive interactions was used to study the observed ferromagnetism~\cite{von}. Mass imbalance provided a clearer signature of ferromagnetic phase. Also, the correlated states of moire lattices which appear at commensurate fillings can be ferromagnetic due to the prsence of quasi flat bands in the system~\cite{Wu}. The modulation in the magnitude of neighboring spins in the observed spin imbalanced ferromagnetism can be compared to mixed spin ferromagnets which have been studied in the past ~\cite{Tucker}.

There also arises a possibility of superconductivity (SC) co-existing with magnetic order in this model in the $U\sim \Delta \gg t_{\sigma},t_{\sigma}'$ limit. We observe a meta-stable singlet d-wave or extended s-wave SC phase in presence of mass imbalance which may become partly stable and continue co-existing with weak magnetic order for low values of mass imbalance. However, an interesting inhomogeneous Fulde-Ferrell-Larkin-
Ovchinnikov (FFLO) state~\cite{FF,LO,Casalbouni,Wang2,Wang3,Muller}  may become a favoured ground state and  can arise specially because of high Zeeman field owing to large uniform magnetization admitted by the system at high values of mass imbalance. Other superfluid phases like breached pair state~\cite{Breach} or Sarma phase~\cite{Sarma,Baarsma} can also arise in this system.

The paper is organized as follows. In sec. II, we discuss our results for low to intermediate values of $U,\Delta$ using unrestricted Hartee-Fock theory. In sec. III, we briefly discuss the limit $U\gg \Delta,t_{\sigma},t_{\sigma}'$ qualitatively. Next in sec. IV, we discuss our results for the limit $U,\Delta \gg t_{\sigma},t_{\sigma}'$ in details. Lastly, in sec. V we conclude.

\section{Hartree-Fock theory}

We first solve the mass imbalanced IHM with frustration in the limit of low to intermediate values of correlation and onsite ionic potential using unrestricted Hartree-Fock (HF) theory. The mass imbalanced IHM is described by the following Hamiltonian,

\begin{align}
    \mathcal{H}=&-\sum_{i,j\sigma}(t_{ij\sigma}c_{i\sigma}^{\dagger}c_{j\sigma} + h.c.)-\mu\sum_{i}n_{i} \nonumber\\
    &-\frac{\Delta}{2}\sum_{i \in A} n_i +\frac{\Delta}{2}\sum_{i \in B} n_i +U\sum_{i}n_{i\uparrow}n_{i\downarrow}
    \label{model}
\end{align}

where $t_{ij\sigma}=t_{\sigma}$ is the hopping amplitude connecting $i,j$ sites which are nearest neighbors while $t_{ij\sigma}=t_{\sigma}'$ is the hopping amplitude connecting $i,j$ sites which are next nearest neighbors. $t_{ij\sigma}=0$ for farther neighbor bonds. Here, $U$ is the onsite Coulomb repulsion and $-\Delta/2$ is the ionic potential on A sublattice where as $\Delta/2$ is the ionic potential on B sublattice. $\mu$ is the chemical potential which fixes the average density to be unity. We define the hopping asymmetry parameter, $\eta=M_{\uparrow}/M_{\downarrow}=t_{\downarrow}/t_{\uparrow}=t_{\downarrow}'/t_{\uparrow}'$, which is defined as the ratio of the down-spin and up-spin nearest or next-nearest neighbor hopping amplitudes and is inversely related to the mass of the fermionic species carrying up and down pseudo-spins. Asymmetry in hopping amplitudes arise because the heavier mass fermion moves slower as compared to the lighter mass fermion. In the HF analysis, we consider $t_{\downarrow}=1$ and $t_{\uparrow}=0.5t_{\downarrow}$. Similarly, $t_{\downarrow}'=0.3t_{\downarrow}$ and $t_{\uparrow}'=0.15t_{\downarrow}$, such that $\eta=2$, unless otherwise mentioned.

Within HF theory we decompose the Coulomb term keeping non zero expectation values for spin resolved densities, $n_{\alpha\sigma}$ where $\alpha \in A,B$, which we solve self-consistently. The details of the calculation are given in Appendix A. From  $n_{\alpha\sigma}$, we construct linear combinations which are defined as the staggered magnetization, $m_{s}=(m_{A}-m_{B})/2$, and the uniform magnetization, $m_{f}=(m_{A}+m_{B})/2$, where $m_{\alpha}=n_{\alpha\uparrow}-n_{\alpha\downarrow}$ is the sublattice magnetization. We also calculate density difference between two sublattices, $\delta=(n_{A}-n_{B})/2$ which is a positive quantity since $A$ sublattice prefers higher particle density than the $B$ sublattice owing to the potential energy wells at $A$ and hills at $B$.

\begin{figure}
    \centering
    \includegraphics[scale=0.245,angle=-90]{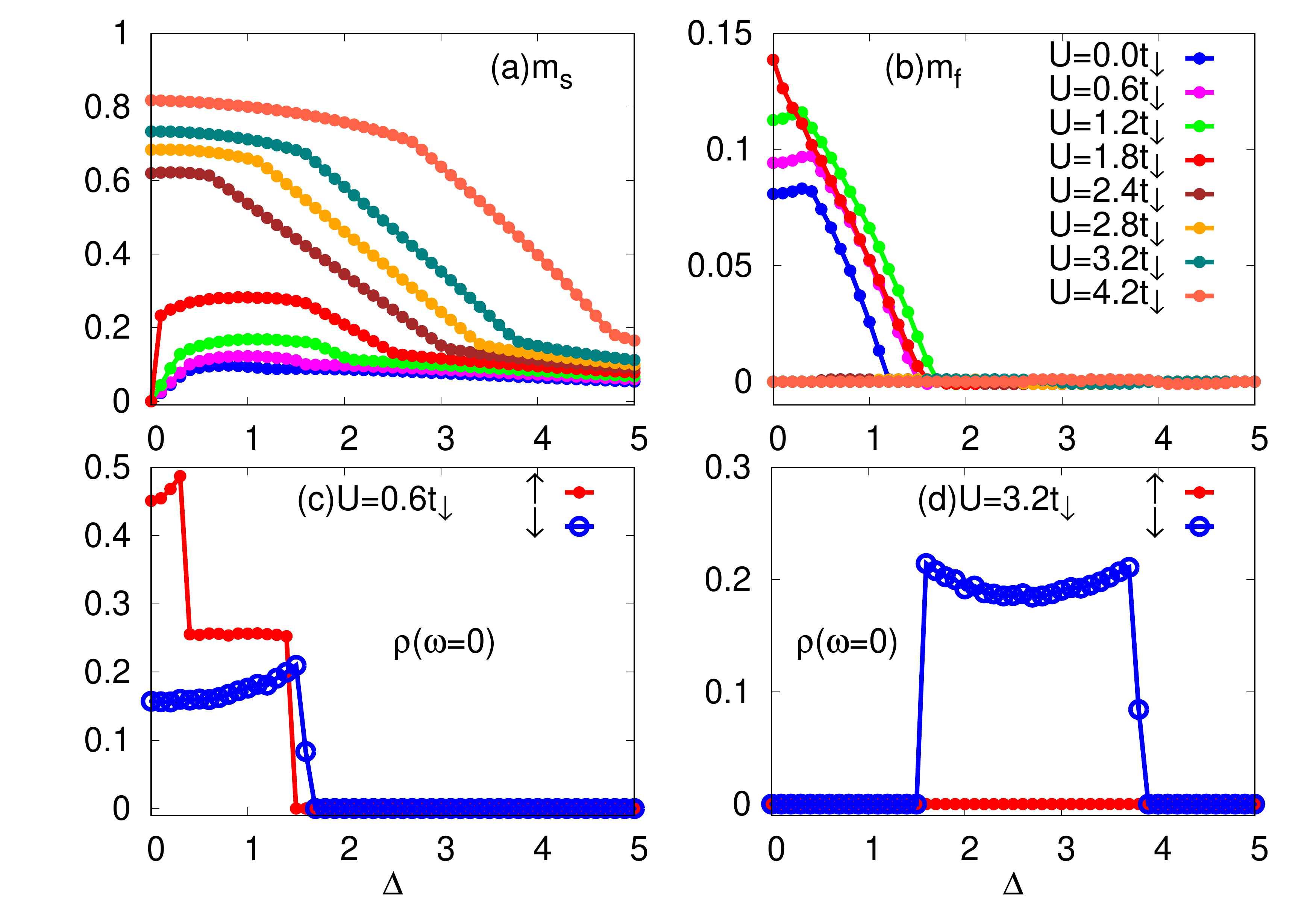}
    \caption{(a)-(b) show staggered magnetization, $m_{s}$ and uniform magnetization, $m_{f}$ as a function of  $\Delta$ for different values of $U$ for $\eta=2$. The nature of $m_{s}$  changes above a threshold value of $U \sim 1.8t_{\downarrow}$ in the sense that at $\Delta=0$, non-zero $m_{s}$ develops only for $U\geqslant1.8t_{\downarrow}$. $m_{f}$ is non-zero over a range of $\Delta$ for $U\leqslant1.8t_{\downarrow}$ which indicates the coexistence of $m_{s}$ and $m_{f}$ both for these parameter values and hence there can be a spin imbalanced ferromagnetic and/or ferrimagnetic phase. (c)-(d) show density of states at fermi level, $\rho(\omega \sim0)$ for $U=0.6t_{\downarrow}$ and $U=3.2t_{\downarrow}$ respectively. (c) tells us that the spin imbalanced ferromagnetic and/or ferrimagnetic phase is metallic and also there exist a down-spin polarized AF half-metallic phase. (d) tells us that between two insulating phases there is a broad phase of down-spin polarized AF half metal. }
    \label{fig:msmfdos}
\end{figure}

In Fig.~\ref{fig:msmfdos}, we show staggered magnetization, $m_{s}$ and uniform magnetization, $m_{f}$ as a function of ionic potential, $\Delta$ for a wide range of $U$ values. For values of $U \leqslant 1.8t_{\downarrow}$, at $\Delta=0$ (asymmetric HM), $m_{s}=0$. Increasing $U$ further, makes $m_{s}$  non-zero at $\Delta=0$. We see that with increase in $U$, $m_{s}$ increases in general and remains non-zero throughout the parameter space because of the explicit breaking of $SU(2)$ symmetry. $m_{f}$ exists for $U \leqslant 1.8t_{\downarrow}$ over a range of $\Delta$ and it increases with increase in $U$ before going to zero for higher $U$ values. This means that for these parameter values $m_{s}$ and $m_{f}$ coexist giving rise to a spin imbalanced ferromagnetic and/or ferrimagnetic phase where spins align parallel/anti-parallel to each other on alternate sites but with unequal magnitudes. The exact nature of the phase is discussed following Fig.~\ref{fig:mAmBnAnB}. The single particle density of states is defined as, $\rho_{\sigma}(\omega)=-1/2\sum_{k,\alpha} \text{Im}G_{\alpha\sigma}(k,\omega^{+})/\pi$ where $G_{\alpha\sigma}(k,\omega)$ is the spin resolved single particle Green's function for $\alpha \in A,B$ sublattice. (c) and (d) show the single particle density of states at the fermi level, $\rho(\omega\sim0)$ as a function of $\Delta$ for $U=0.6t_{\downarrow}$ and $U=3.2t_{\downarrow}$ respectively. We find that the spin imbalanced ferromagnetic and/or ferrimagnetic phase is metallic having finite density of states in both spin channels such that $\rho_{\uparrow}(\omega \sim0) \neq \rho_{\downarrow}(\omega\sim0) $. For $U=0.6t_{\downarrow}$, the spin imbalanced ferro-/ferri-magnetic phase leads to a down spin polarized AF half metallic phase before it goes to an insulating phase upon increasing $\Delta$. For $U=3.2t_{\downarrow}$, the broad down spin polarized AF half-metallic phase is bounded by insulating phases on both sides. It is understandable that due to the heavier up spin species and mobile down spin species, the AF half-metallic phases are down spin polarized. For higher values of $U$, the constant plateau regions at low and high values of $\Delta$ in the $m_{s}$ curves indicates insulating phases.

\begin{figure}
    \centering
    \includegraphics[scale=0.245,angle=-90]{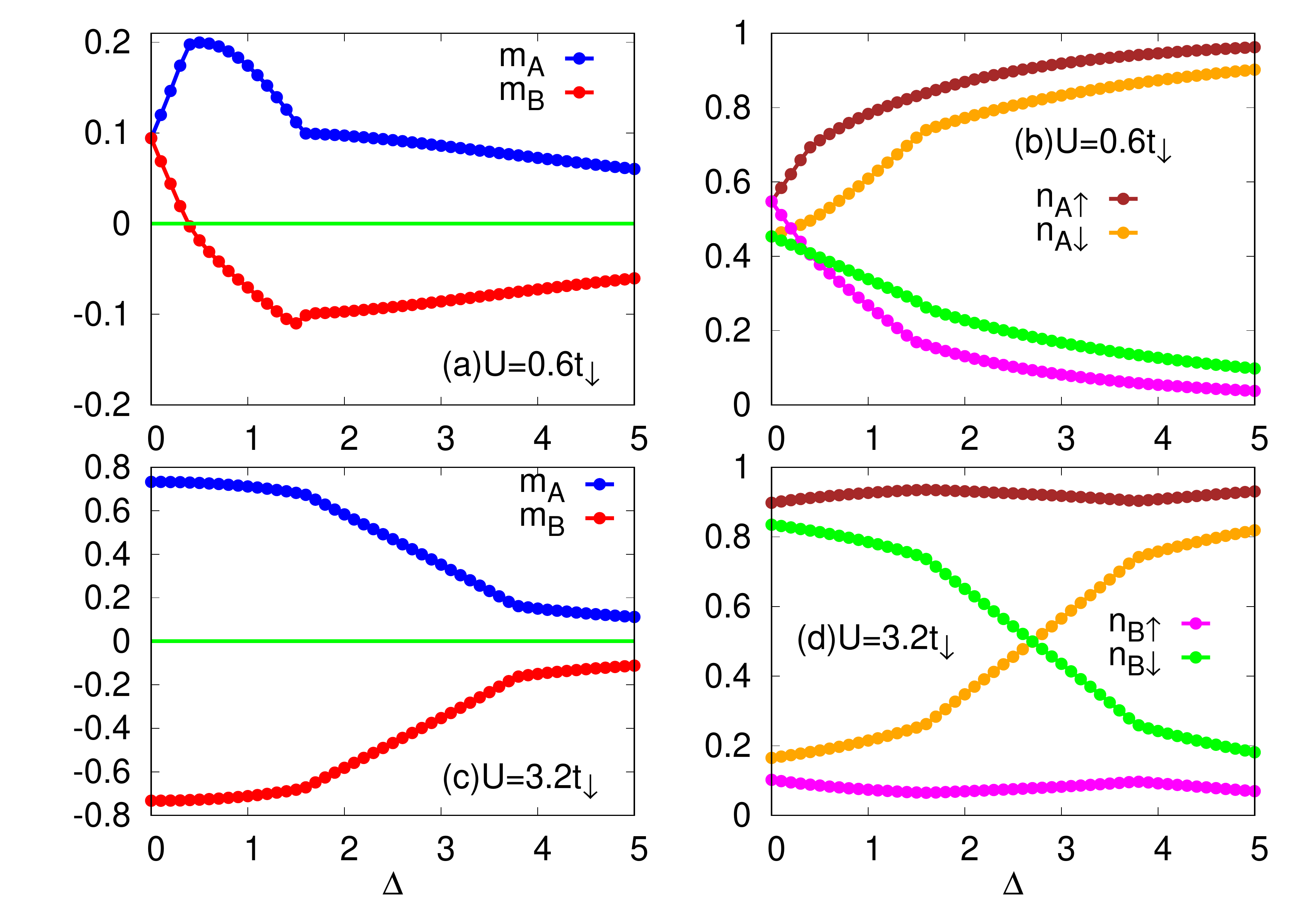}
    \caption{(a) and (c) give the plot of individual sublattice magnetizations for $U=0.6t_{\downarrow}$ and U=$3.2t_{\downarrow}$ respectively. From (a) we find that there is a region of parameter space for which $m_{A},m_{B}>0$ but $m_{A} \neq m_{B}$ (apart from $\Delta=0$ where $m_{A}=m_{B}$). Thus, in general this phase is a spin-imbalanced ferromagnetic phase. There is also a regime of $\Delta$ where $m_{A}>0,m_{B}<0$ and $m_{A} \neq |m_{B}|$. This is a ferrimagnetic phase and as shown earlier both phases are metallic in nature. For $U=3.2t_{\downarrow}$ such phases are absent. (b) and (d) give the spin resolved densities, $n_{\alpha\sigma}$ as a function of $\Delta$. It shows that magnetic order, $m_{\alpha}=(n_{\alpha\uparrow}-n_{\alpha\downarrow})$ exists for all parameter values and also $n_{A}\geqslant n_{B}$ signifying positive density difference between the sublattices.  }
    \label{fig:mAmBnAnB}
\end{figure}

Fig.~\ref{fig:mAmBnAnB} shows the plots of sublattice magnetization, $m_{\alpha}$ and the spin resolved densities, $n_{\alpha\sigma}$ for $U=0.6t_{\downarrow}$ and $U=3.2t_{\downarrow}$. For $U=0.6t_{\downarrow}$, $m_{A},m_{B}>0$ but $m_{A} \neq m_{B}$ (at $\Delta=0$, $m_{A}=m_{B}$) for a range of $\Delta$ which means this a spin imbalanced ferromagnetic phase.  The spin imbalanced ferromagnetic phase leads to a regime where $m_{A}>0,m_{B}<0$ but $m_{A} \neq |m_{B}|$. This is basically a ferrimagnetic phase. Both these phases are metallic in nature as discussed earlier. Beyond the ferrimagnetic metallic phase, $m_{A}=-m_{B}$ such that $m_{f}=0$ and the phase is insulating. For $U=3.2t_{\downarrow}$, $m_{A}=-m_{B}$ for the entire parameter space. If we now look at the spin resolved densities for $U=0.6t_{\downarrow}$, we find that at $\Delta=0$, $n_{A\sigma}=n_{B\sigma}$ which makes $\delta=0$ in contrast to $U=3.2t_{\downarrow}$ case where we find $\delta \neq 0$ even at zero ionicity. This is purely a mass imbalance effect which can be seen from the self-consistent equation of $\delta=-1/N^{2}\sum_{k,\sigma}\tilde{\Delta}_{\sigma}/\sqrt{{\tilde{\Delta}_{\sigma}}^2+4t_{\sigma}^2(\cos(k_{x})+\cos(k_{y}))^{2}}$, where $\tilde{\Delta}_{\sigma}=U(\delta-\sigma m_{s})/2$ is the effective ionic potential that the system feels for $\Delta=0$. This equation is valid in the insulating phase for larger values of U at zero ionicity. We see that the asymmetry in the hopping parameter is the reason behind a weak density difference at $\Delta=0$ for larger values of $U$. We also see that $n_{A}\geqslant n_{B}$ for all parameter values which means due to potential wells at A sites, particle density is more on A sublattice than on B sublattice. We now comment about the nature of insulating phases. It is seen that both charge order and spin order exists for the insulating phases. However, when $\Delta$ is low, the gap in the spectrum is predominantly due to spin order in the system where as when $\Delta$ is high the gap is predominantly due to charge order prevailing in the system. We call the insulating phase at low values of $\Delta$, an AF insulating phase (though it has weak charge modulation) and the insulating phase at high values of $\Delta$, a correlated band insulator with weak AF spin density wave (SDW) order. Also, we observe from the plot of spin resolved densities that the probability of occupancy of the up spin species on A sublattice is more compared to the down spin species which is reverse for the B sublattice in most of the parameter space. This is particularly striking in the higher ionicity region where spontaneously induced magnetization does not exist and magnetization is mass imbalance induced. This is because the heavier mass up spin species prefers to sit at the potential wells as compared to the lighter mass down spin species.

\begin{figure}
    \centering
    \includegraphics[scale=0.32,angle=-90]{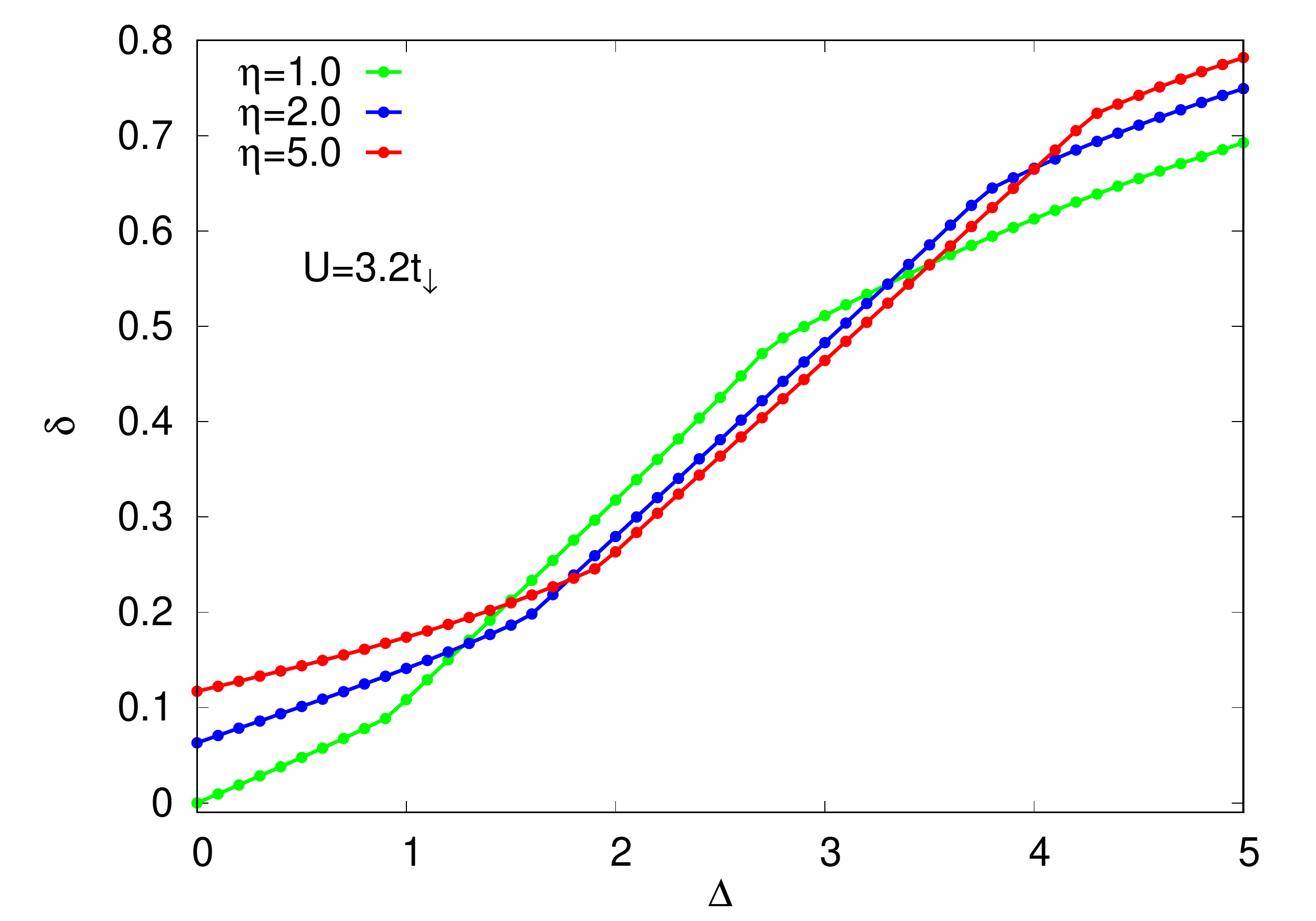}
    \caption{ Figure shows the density difference between sublattices for different $\eta$ values for $U=3.2t_{\downarrow}$. The small density difference at zero ionicity is a pure mass imbalance effect which is enhanced with increasing asymmetry in hopping. Also, density difference in general increases with ionicity and charge order prevails in the entire parameter space. }
    \label{fig:dendiff}
\end{figure}

Fig.~\ref{fig:dendiff} reiterates the fact that the non zero density difference between the sublattices at zero ionicity, for relatively larger values of $U$,  is a purely mass imbalance induced effect. The density difference in general increases with ionicity for any $\eta$ but is larger for increased hopping asymmetry in the insulating phases. Charge order exists throughout the parameter regime.

\begin{figure}
    \centering
    \includegraphics[scale=0.32,angle=-90]{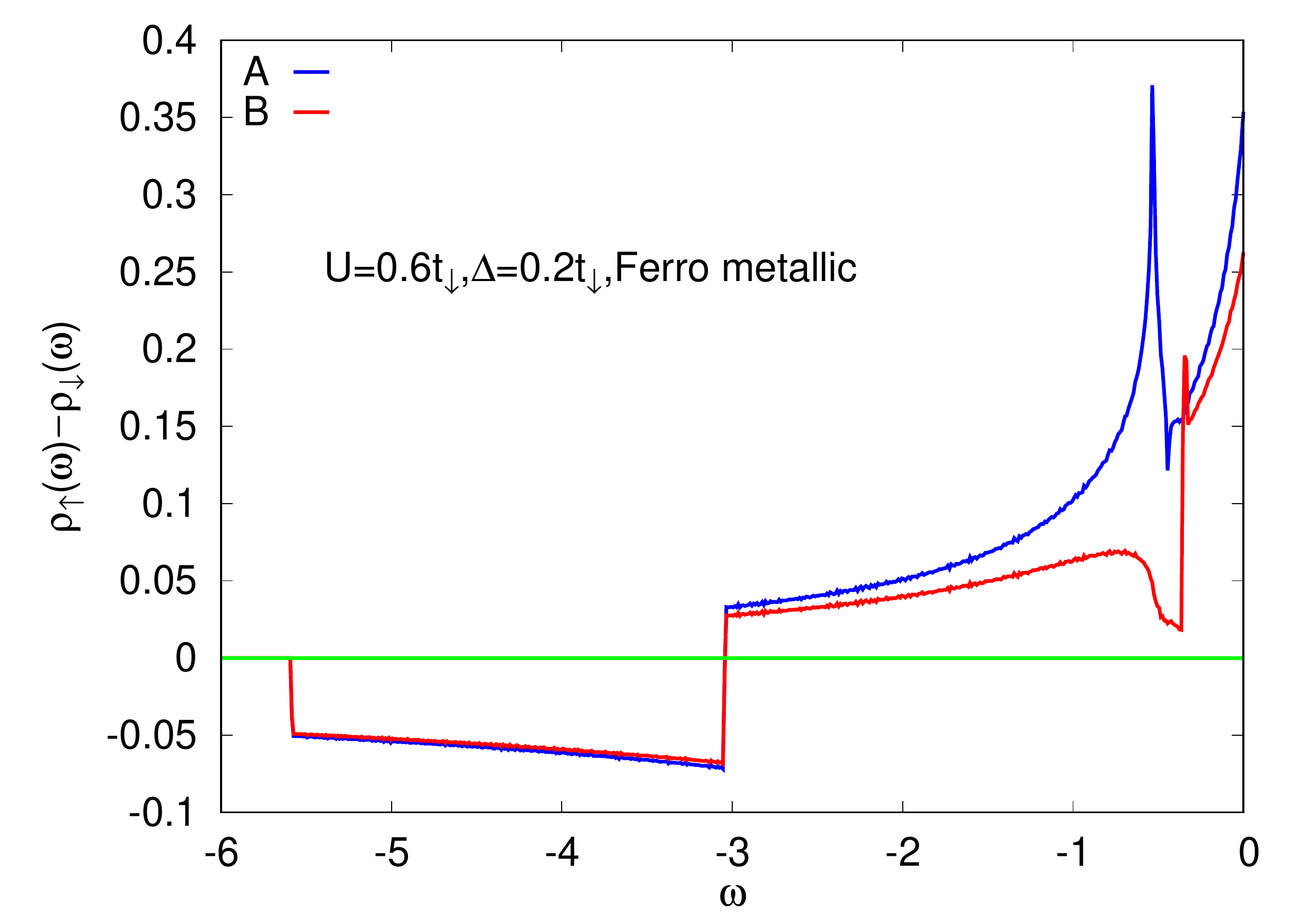}
    \caption{ Figure shows difference in the single particle density of states in up and down spin channels, $\rho_{\uparrow}(\omega)-\rho_{\downarrow}(\omega)$ as a function of $\omega$ in the occupied part of the spectrum at half-filling for $U=0.6t_{\downarrow},\Delta=0.2t_{\downarrow}$. In this ferromagnetic phase, if we integrate over the occupied spectrum ,it is clear that $m_{A},m_{B}>0$ and $m_{A}>m_{B}$. }
    \label{fig:updown}
\end{figure}

Fig.~\ref{fig:updown} shows the difference in the single particle density of states between the up and down spin channels, $\rho_{\uparrow}(\omega)-\rho_{\downarrow}(\omega)$ with $\omega$ below the chemical potential in the ferromagnetic phase. Due to higher positive weight for both A and B sublattices, if we intregate this quantity over the occupied spectrum, $m_{A},m_{B}$ will both be positive. Also it is clear that $m_{A}>m_{B}$  in this phase.

\begin{figure}
    \centering
    \includegraphics[scale=0.35,angle=-90]{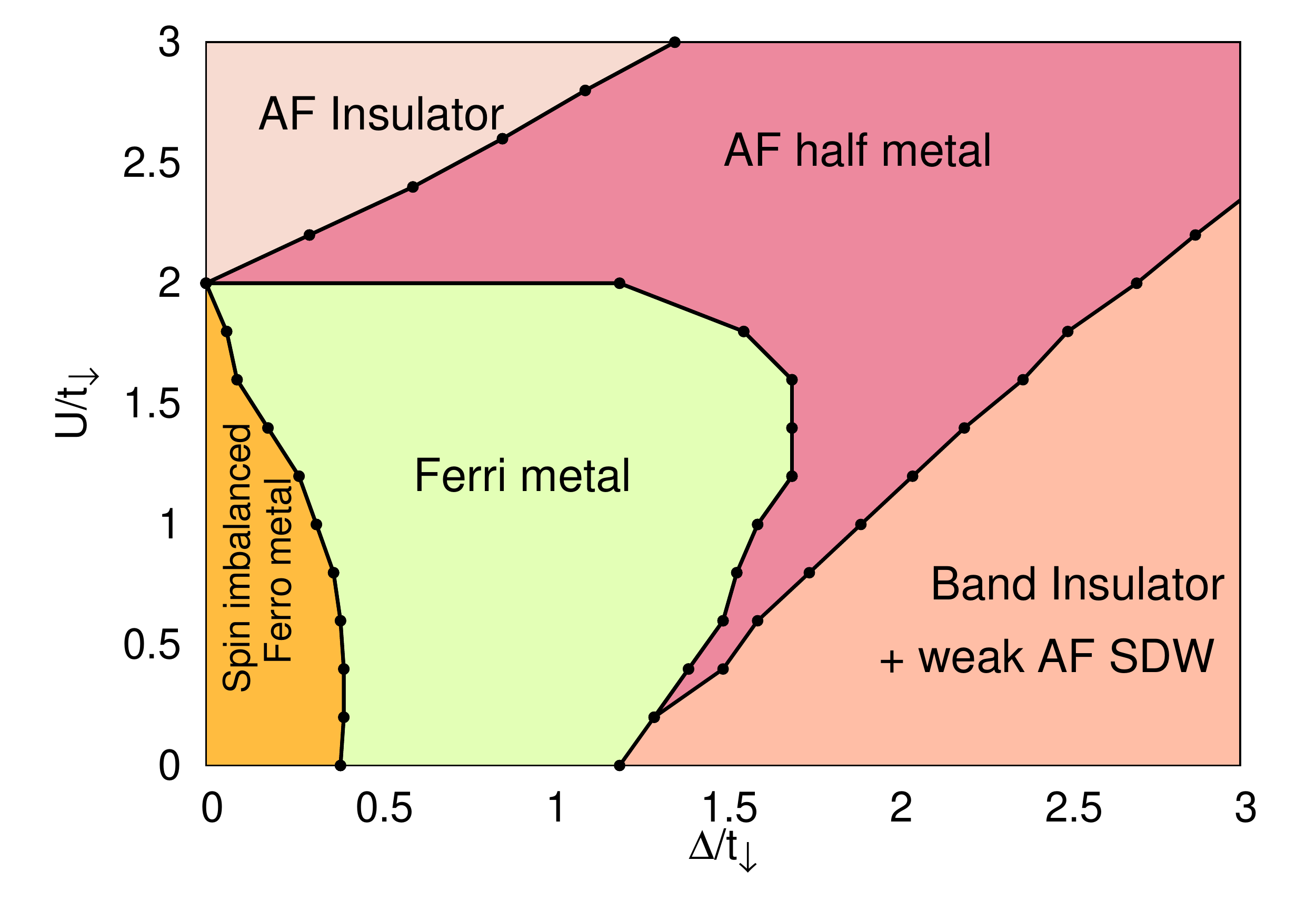}
    \caption{ Figure shows phase diagram constructed using unrestricted HF theory for $\eta=2$ in the $U-\Delta$ plane. Between the AF insulating phase and the band insulator phase with weak AF SDW order, there is a broad range of magnetically ordered metals like spin imbalanced ferromagnetic metal, ferrimagnetic metal and AF half metal.  }
    \label{fig:hfpd}
\end{figure}

\begin{figure*}
    \centering
    \includegraphics[scale=0.35,angle=-90]{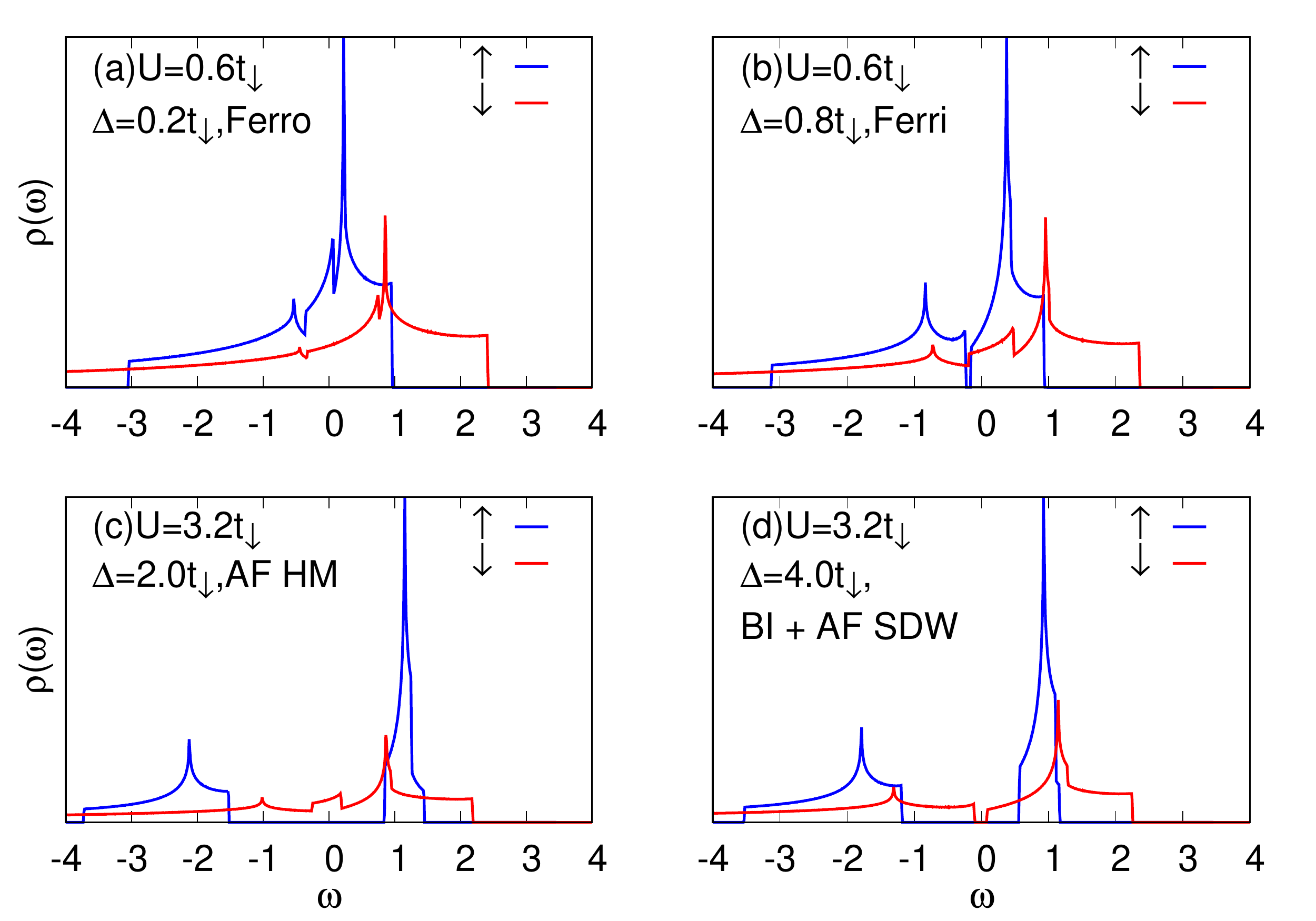}\includegraphics[height=8.5  cm,angle=-90]{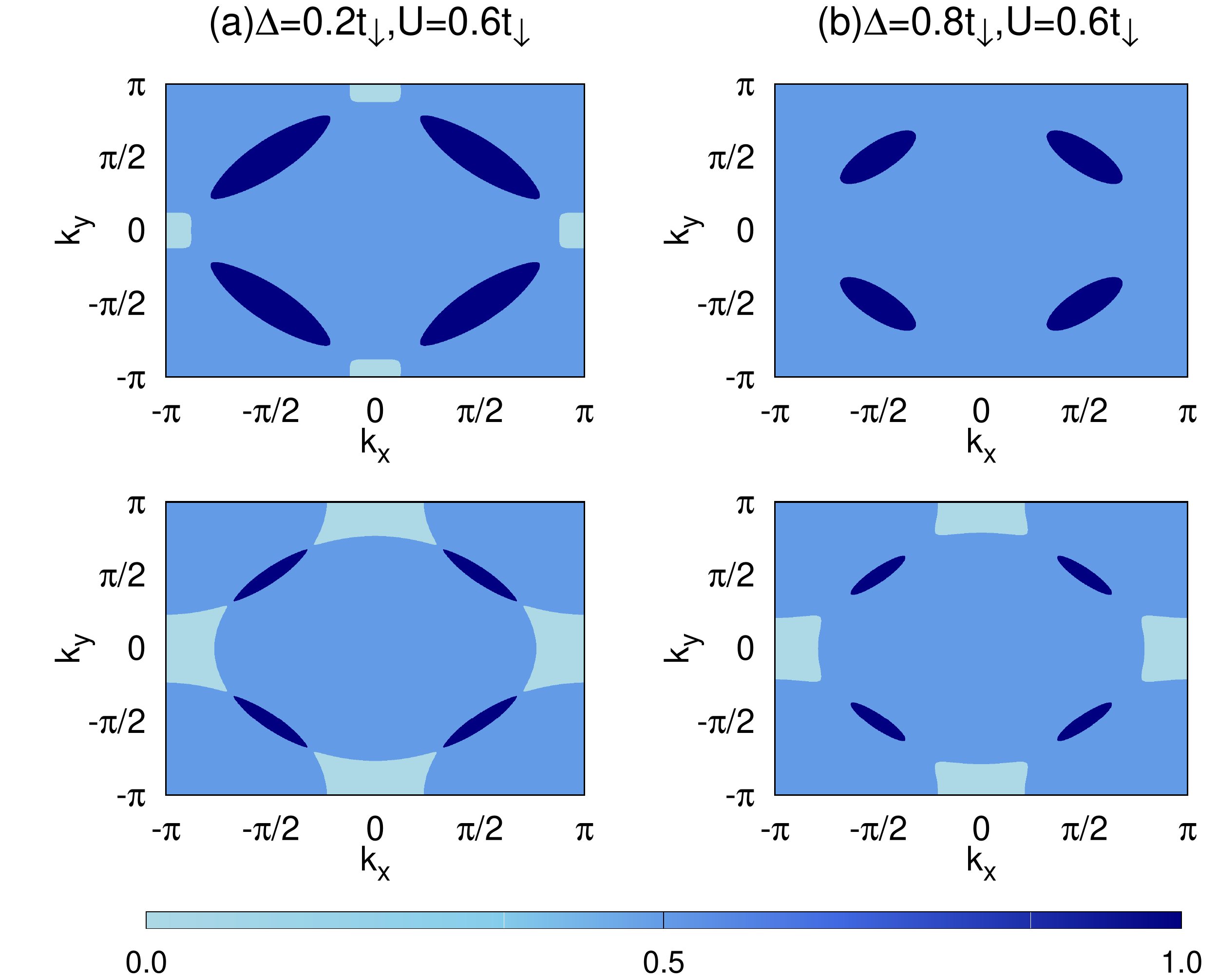}
    \caption{Left: Figure shows density of states, $\rho(\omega)$ versus $\omega$ for the (a) ferromagnetic metallic phase (b) ferrimagnetic metallic phase (c) AF half metallic phase and (d) band insulator with weak SDW order phase. For both ferromagnetic and ferrimagnetic phases, there is finite density of states at $\omega \sim 0$ making these phases metallic. The AF half metallic phase has down spin polarized conductivity with a gap in the up spin channel at the fermi level. The band insulating phase which has weak AF SDW order has gap in both spin channels at the fermi level. Right: Figure shows momentum distribution function, $n_{k\sigma}$ in the full Brillouin zone for the ferromagnetic and ferrimagnetic metallic phases. The upper row is for the up spin polarity and the lower row is for the down spin polarity. In the ferro phase, there are electron ($n_{k\sigma}>1/2$) and hole ($n_{k\sigma}<1/2$) pockets in both spin channels with larger electron pockets in the up spin channel and larger hole pockets in the down spin channel. In the ferri phase, the up spin channel has only electron pockets where as down spin channel has both electron and hole pockets. }
    \label{fig:hfdos}
\end{figure*}

The phase diagram in the $U-\Delta$ plane is shown in Fig.~\ref{fig:hfpd} for $\eta=2$. When $\Delta=0$, the asymmetric HM shows a metal-insulator transition from a ferromagnetic metallic phase to AF insulating phase with increasing $U$. For $U=0$, the system again shows a metal-insulator transition with increasing $\Delta$. This non-interacting limit is exactly solvable with eigen energies, $\lambda^{1,2}_{k\sigma}=-\mu-4t_{\sigma}'\cos(k_{x})\cos(k_{y})\mp \sqrt{4t_{\sigma}^{2}(\cos(k_{x})+\cos(k_{y}))^{2}+\Delta^{2}/4}$. The conduction band  minima occurs at $(k_{x},k_{y})=(\pm \pi/2,\pm \pi/2)$ with $\lambda^{2,min}_{\sigma}= \Delta/2$ for bands of both spin polarity. Where as, the valence band maxima occur at $(k_{x},k_{y})=(0,\pm \pi),(\pm \pi,0)$ with
$\lambda^{1,max}_{\sigma}=4t_{\sigma}'- \Delta/2$. Approaching from the band insulating side, the band gap between the bands of down spin species closes first at $\Delta=1.2t_{\downarrow}$ (taking $t_{\downarrow}'=0.3t_{\downarrow})$ where there is a metal-insulator transition. But since, the conduction band minima corresponds to $\Delta/2$ for both spin polarity bands, for $\Delta\leqslant1.2t_{\downarrow}$, the conduction band of the up-spin species also crosses the fermi level. Therefore both spin species show metallicity simultaneously at half-filling, whence we get a ferrimagnetic metallic phase. On further lowering $\Delta$, we get a spin imbalanced ferromagnetic phase. When $U,\Delta$ are both non-zero, we get a broad intermediate range of $SU(2)$ symmetry broken metals consisting of an AF half metal, a novel spin imbalanced ferromagnetic metal and a significantly broader range of ferrimagnetic metal than observed in an earlier study of frustrated IHM without mass imbalance~\cite{Bag2}, between two insulating phases.  

Introducing mass imbalance effectively reduces the bandwidth of the heavier up spin species as compared to the down spin species, as is visible in the left panel of Fig.~\ref{fig:hfdos} (a)-(d). This means effectively up spin polarized band gets flatter resulting in high valued peaks in the density of states. In (a), $\rho_{\uparrow}(\omega \sim 0)$ is relatively large due to the relatively flatter band ($\eta=2$). As mentioned earlier existence of such quasi flat bands can result into ferromagnetism. In this ferromagnetic phase, the fermi level lies in the region where both conduction and valence bands overlap and contribute to the density of states at $\omega \sim 0$ in both up and down spin channel. Consequently, we see in (a) of right panel that the momentum distribution function, defined as $n_{k\sigma}=\int_{-\infty}^{0} A_{k\sigma}(\omega) d\omega=(n_{kA\sigma}+n_{kB\sigma})/2$, where $A_{k\sigma}(\omega)$ is the spin resolved single particle spectral function, has both electron ($n_{k\sigma}>1/2$) and hole ($n_{k\sigma}<1/2$) pockets in both spin channels. The electron pockets are larger in the up spin channel and hole pockets are larger in the down spin channel. An electron pocket arises when the conduction band minima lies below the fermi level such that it is partially occupied where as a hole pocket arises when the valence band maxima is above the fermi level and becomes partially unoccupied. Since, $m_{f}=1/N^2\sum_{k \in FBZ} (n_{k\uparrow}-n_{k\downarrow})$, the unequal sizes of the electron, hole pockets in the two spin channels make $\sum_{k \in FBZ} n_{k\uparrow}>\sum_{k \in FBZ}n_{k\downarrow}$ and thus $m_{f}>0$. In the ferrimagnetic phase, there is finite $\rho_{\sigma}(\omega \sim 0)$ for both spin channels (left panel,(b)) and correspondingly in (b) of the right panel, the up spin momentum distribution function has only electron pockets (fermi level lies within conduction band) where as the down spin channel has both electron and hole pockets ( both valence and conduction band cross the fermi level). Left panel (c) shows that in the AF half metallic state the up spin channel is gapped where as $\rho_{\downarrow}(\omega=0)\neq 0$. (d) shows that in the band insulating phase with weak AF SDW order, both spin channels are gapped at the fermi level. However, in all these cases $\rho_{\uparrow}(\omega) \neq \rho_{\downarrow}(\omega)$. 

In the following sections we will discuss other limits of this model in detail.

\section{$U\gg \Delta,t_{\sigma},t_{\sigma}'$ limit}

In the limit $U\gg \Delta,t_{\sigma},t_{\sigma}'$, double occupancies are energetically expensive on both A and B sublattices and should be eliminated from the low energy Hilbert space. For this, we do a similarity transformation, $\mathcal{H}_{eff}=e^{-iS}\mathcal{H} e^{iS}$ where, $S=-\frac{i}{U}\sum_{\alpha\in A,B}({H_{t'}^{+}}_{\alpha\rightarrow\alpha}-{H_{t'}^{-}}_{\alpha\rightarrow\alpha})-\frac{i}{U+\Delta}({H_{t}^{+}}_{A\rightarrow B}-{H_{t}^{-}}_{ B \rightarrow A})-\frac{i}{U-\Delta}({H_{t}^{+}}_{B\rightarrow A}-{H_{t}^{-}}_{ A \rightarrow B})$. ${H_{t/t'}^{+}}$ represent hopping processes which increase the number of double occupancies and holes by one and ${H_{t/t'}^{-}}$ represent hopping processes which decrease the number of double occupancies and holes by one. The effective low energy Hamiltonian at half-filling in this limit is given by,

\begin{align}
    \mathcal{H}=&\sum_{<ij>}\dfrac{4t_{\uparrow}t_{\downarrow}U}{U^{2}-\Delta^{2}}(S_{iA}^{x}S_{jB}^{x}+S_{iA}^{y}S_{jB}^{y})+\dfrac{2U(t_{\uparrow}^{2}+t_{\downarrow}^{2})}{U^{2}-\Delta^{2}}\nonumber\\&\bigg(S_{iA}^{z}S_{jB}^{z}-\dfrac{1}{4}\bigg)
    +\sum_{i}\dfrac{4\Delta(t_{\uparrow}^{2}-t_{\downarrow}^{2})}{U^{2}-\Delta^{2}}(S_{iA}^{z}-S_{jB}^{z})\nonumber\\&+\sum_{\substack{<<ij>>,\\\alpha \in A,B}}\dfrac{4t'_{\uparrow}t'_{\downarrow}}{U}(S_{i\alpha}^{x}S_{j\alpha}^{x}+S_{i\alpha}^{y}S_{j\alpha}^{y})\nonumber
    +\dfrac{2({t'}_{\uparrow}^{2}+{t'}_{\downarrow}^{2})}{U}\nonumber\\&\bigg( S_{i\alpha}^{z}S_{j\alpha}^{z}-\dfrac{1}{4}\bigg)
\end{align}

Here, the kinetic energy term is completely projected out due to the half-filling constraint which means even holes are not allowed in the low energy Hilbert space in addition to doublons. There exists two competing asymmetric Heisenberg terms on the nearest neighbor and next nearest neighbor bonds which tries to create singlet pairs on the nearest or next nearest neighbor bonds respectively. In addition to these, there is a purely mass imbalance induced staggered magnetic field term which is $\propto (t_{\uparrow}^{2}-t_{\downarrow}^{2})$. Due to the singly occupied frozen state in this limit, no metallic phases are possible. We now look at a more interesting limit which admits metallic phases in the following section.

\section{$U \sim \Delta \gg t_{\sigma},t_{\sigma}'$ limit}

The limit $U \sim \Delta \gg t_{\sigma},t_{\sigma}'$ is interesting in the sense that even in this strongly correlated limit finite nearest and next nearest neighbor hoppings survive in the system at half-filling. Even though large $U$ and $\Delta$ independently promote insulating tendencies e.g,, the Mott insulator and the band insultor respectively, when they are simultaneously present, they compete with each other and can open up a plethora of charge dynamical phases with the help of low energy hoppings. In this limit at half-filling, holes are energetically expensive on A sublattice where the ionic potential is $-\Delta/2$ and doublons are expensive on B sublattice where the ionic potential is $\Delta/2$. We do a similarity transformation, $\mathcal{H}_{eff}=e^{-iS}\mathcal{H} e^{iS}$ which does a site selective projection of doublons or holes. The similarity operator in this case is $S=-\frac{i}{U}\sum_{\alpha\in A,B}({H_{t'}^{+}}_{\alpha\rightarrow\alpha}-{H_{t'}^{-}}_{\alpha\rightarrow\alpha})-\frac{i}{U+\Delta}({H_{t}^{+}}_{A\rightarrow B}-{H_{t}^{-}}_{ B \rightarrow A})-\frac{i}{\Delta}({H_{t}^{0}}_{A\rightarrow B}-{H_{t}^{0}}_{ B \rightarrow A})$. ${H_{t}^{0}}$ are hopping processes which do not change the number of double occupancies or holes in the system. Interestingly, the low energy Hilbert space allows hopping processes ${H_{t}^{+}}_{B\rightarrow A}$ and it's conjugate process ${H_{t}^{-}}_{A\rightarrow B}$ where we can create a doublon on A sublattice and a hole on B sublattice from single occupancies and vice-versa. The low energy effective Hamiltonian is separated as $H_{0}$, the unperturbed part of the Hamiltonian which is basically the energy of the doublons on A sublattice and holes on B sublattice, $H_{hopp}$, the kinetic energy terms, $H_{d}$, the dimer terms and $H_{tr}$, the trimer terms. All these terms are in the projected space where holes are not allowed on A sublattice and doublons are not allowed on B sublattice represented by the projection operator, $\mathcal{P}=\prod_{i,j}(1-(1-n_{iA\uparrow})(1-n_{iA\downarrow}))(1-n_{jB\uparrow}n_{jB\downarrow})$.
\begin{align}
    H_{0}=\sum_{i}\dfrac{U-\Delta}{2}[n_{iA\uparrow}n_{iA\downarrow}+(1-n_{iB\uparrow})(1-n_{iB\downarrow})]
    \end{align}
    
    \begin{align}
    H_{hopp}=&-\sum_{<ij>,\sigma}t_{\sigma}(c_{iA\sigma}^{\dagger}c_{jB\sigma} +h.c.)\nonumber\\&-\sum_{\substack{<<ij>>,\\\sigma,\alpha}}t'_{\sigma}(c_{i\alpha\sigma}^{\dagger}c_{j\alpha\sigma} +h.c.)
\end{align}

\begin{align}
    H_{d}=&\sum_{<ij>}\dfrac{2t_{\uparrow}t_{\downarrow}}{U+\Delta}(S_{iA}^{x}S_{jB}^{x}+S_{iA}^{y}S_{jB}^{y})+\dfrac{(t_{\uparrow}^{2}+t_{\downarrow}^{2})}{U+\Delta}\nonumber\\&\bigg(S_{iA}^{z}S_{jB}^{z}-\dfrac{(2-n_{iA})n_{jB}}{4}\bigg)+\dfrac{(t_{\uparrow}^{2}-t_{\downarrow}^{2})}{U+\Delta}\nonumber\\&\bigg(\dfrac{2-n_{iA}}{2}S_{jB}^{z}-\dfrac{n_{jB}}{2}S_{iA}^{z}\bigg)+\sum_{\substack{<<ij>>,\\\alpha}}\bigg[\dfrac{4{t'}_{\uparrow}{t'}_{\downarrow}}{U}\nonumber\\&(S_{i\alpha}^{x}S_{j\alpha}^{x}+S_{i\alpha}^{y}S_{j\alpha}^{y})+\dfrac{2({t'}_{\uparrow}^{2}+{t'}_{\downarrow}^{2})}{U} S_{i\alpha}^{z}S_{j\alpha}^{z}\bigg]\nonumber\\&-\dfrac{2({t'}_{\uparrow}^{2}+{t'}_{\downarrow}^{2})}{U}\sum_{<<ij>>}\bigg(\dfrac{(2-n_{iA})(2-n_{jA})}{4}\nonumber\\&+\dfrac{n_{iB}n_{jB}}{4}\bigg)
 -\sum\limits_{<ij>,\sigma}\frac{t_{\sigma}^2}{\Delta}[(1-n_{iA\bar{\sigma}})(1-n_{jB})\nonumber\\&+(n_{iA}-1)n_{jB\bar{\sigma}}]
 \end{align}
 \begin{align}
     &H_{tr}=-\sum_{\substack{<ijk>,\\\sigma}}\dfrac{t_{\sigma}^2}{\Delta}\bigg[c_{kA\sigma}^{\dagger}n_{jB\bar{\sigma}}c_{iA\sigma}+c_{iB\sigma}(1-n_{jA\bar{\sigma}})c_{kB\sigma}^{\dagger}\bigg]\nonumber\\&-\dfrac{t_{\uparrow}t_{\downarrow}}{\Delta}\sum_{\substack{<ijk>,\\\sigma}}\bigg[c_{iA\bar{\sigma}}c_{jB\bar{\sigma}}^{\dagger}c_{jB\sigma}c_{kA\sigma}^{\dagger} +c_{iB\sigma}c_{jA\sigma}^{\dagger}c_{jA\bar{\sigma}}c_{kB\bar{\sigma}}^{\dagger}\bigg]\nonumber\\
    &+\sum_{\substack{<kj>,\\<<ik>>\sigma}}\dfrac{t_{\sigma}{t'}_{\sigma}(U+\Delta)}{2U\Delta}\bigg[c_{iA\sigma}^{\dagger}(1-n_{kA\bar{\sigma}})c_{jB\sigma}\nonumber\\&-c_{jA\sigma}^{\dagger}n_{kB\bar{\sigma}}c_{iB\sigma}\bigg]+\sum_{\substack{<kj>,\\<<ik>>\sigma}}\dfrac{t_{\sigma}{t'}_{\bar{\sigma}}(U+\Delta)}{2U\Delta}\nonumber\\&\bigg[c_{iA\bar{\sigma}}^{\dagger}c_{kA\sigma}^{\dagger}c_{kA\bar{\sigma}}c_{jB\sigma}+c_{jA\sigma}^{\dagger}c_{kB\bar{\sigma}}^{\dagger}c_{kB\sigma}c_{iB\bar{\sigma}}\bigg]+h.c.
\end{align}
 
The detailed derivation of these terms in the $SU(2)$ symmetric case can be found in~\cite{Anwesha1}, co-authored by us. However, the derivation of the explicit mass imbalance induced staggered magnetic field term is shown in Appendix B. Now in order to solve this effective Hamiltonian in the unprojected space, we renormalize the couplings with appropriate statistical weight factors which take into account the site dependent projection of holes or doublons approximately and are known as Gutzwiller factors. Detailed derivation of Gutzwiller factors can be found in \cite{Anwesha1}, co-authored by us. However, we give a list of all the Gutzwiller factors used in our calculation in Appendix B. After obtaining the renormalized Hamiltonian, we solve it using mean field theory keeping the following mean fields in our calculation :(a) staggered and uniform magnetization, $m_{s}=(m_{A}-m_{B})/2,$ $m_{f}=(m_{A}+m_{B})/2$ where $m_{\alpha},\alpha \in A,B$ is the sublattice magnetization (b) density difference between two sublattices, $\delta=(n_{A}-n_{B})/2$ (c) intra sublattice fock shift on A(B) sublattice, with $\chi_{\alpha\alpha\sigma}=\langle c_{i\alpha\sigma}^{\dagger}c_{i\pm 2x/2y \alpha\sigma} $+h.c.$\rangle$ and $\chi_{\alpha\alpha\sigma}^{'}=\langle c_{i\alpha\sigma}^{\dagger}c_{i\pm x \pm y \alpha\sigma}$+h.c.$\rangle$ and (d) inter sublattice fock shifts, $\chi_{AB\sigma}^{(1)}=\langle c_{iA\sigma}^{\dagger}c_{jB\sigma}\rangle,j=i\pm x,i \pm y,\chi_{AB\sigma}^{(2)}=\langle c_{iA\sigma}^{\dagger}c_{jB\sigma}\rangle,j=i\pm 2x\pm y$ or $i \pm 2y \pm x$. For studying singlet pairing, we do a two step Bogoliubov-deGennes (BdG) calculation where inter-band pairing has been considered to be weak, keeping singlet pairing amplitude, $\Delta_{AB}=\langle c_{iA\uparrow}^{\dagger}c_{j B\downarrow}^{\dagger}-c_{iA\downarrow}^{\dagger}c_{j B\uparrow}^{\dagger}\rangle$ as a mean field parameter. Here (i,j) are nearest neighbor sites. We consider two pairing channels : d-wave and extended s-wave. For d-wave, $\Delta_{d}^{\pm y}=-\Delta_{d}^{\pm x}$ and for extended s-wave,  $\Delta_{s}^{\pm y}=\Delta_{s}^{\pm x}$ where, $\pm x$ and $\pm y$ denote nearest neighbors in positive or negative x,y directions.  The above mean fields are then solved self-consistently. We present the results of this renormalized mean field theory for $t_{\uparrow}=1,t_{\uparrow}'=0.3t_{\uparrow}$ and varying $t_{\downarrow},t_{\downarrow}'$ for $U=10t_{\uparrow}$.

\begin{figure}
    \centering
    \includegraphics[scale=0.245,angle=-90]{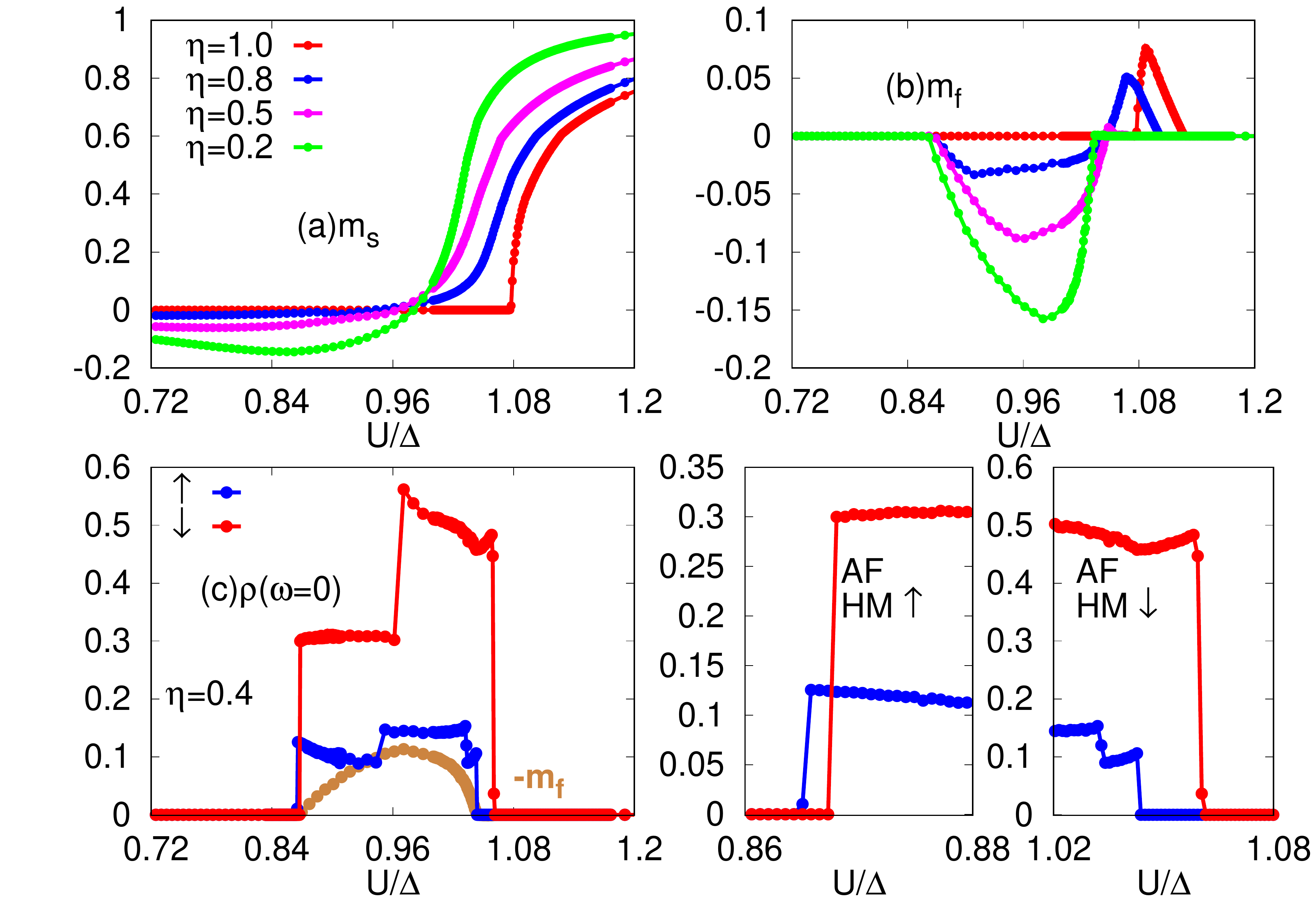}
    \caption{(a) and (b) show the staggered magnetization, $m_{s}$ and uniform magnetization, $m_{f}$ for different values of $\eta$ for $U=10t_{\uparrow}$.  $m_{s}$ flips in sign as $U/\Delta$ is decreased for finite mass imbalance. $m_{f}$ remains positive for a small regime (which diminishes with incresing mass imbalance) and remains negative in most of the non-zero regime for finite mass imbalance. (c) shows the density of states at fermi level, $\rho(\omega \sim 0)$ for $\eta=0.4$ which shows that between two opposite polarity AF half-metals, there is a regime which can be a spin imbalanced ferromagnetic and/or ferrimagnetic phase. The offset shows the AF half-metallic phases more clearly. These metallic phases are bounded by insulating phases on both side.  }
    \label{fig:msmflargeUDelta}
\end{figure}

In Fig.~\ref{fig:msmflargeUDelta}(a)-(b) , we plot staggered magnetization ,$m_{s}$ and uniform magnetization, $m_{f}$ for different values of $\eta$ for $U=10t_{\uparrow}$. $\eta=1$ i.e., the mass balanced case shows a first order transition from an AF ordered state to a PM state where $m_{s}$ goes to zero. For finite $\eta$ however $m_{s}$ remains non-zero throughout the parameter space due to breaking of $SU(2)$ symmetry. For high values of $U/\Delta$, $m_{s}$ increases with increasing hopping asymmetry which is because of the introduction of easy axis anisotropy in the system which helps in
stabilizing N{\'e}el type order. As we decrease $U/\Delta$, $m_{s}$ flips in sign and again the magnitude of negative magnetization increases with asymmetry in hopping. At small values of $U/\Delta$, the potential wells are deep and the heavier mass down spin polarized fermionic species prefers to sit at the potential wells i.e., A sites rather than B sites where the lighter mass up spin polarized fermions live. This makes $m_{A}<0,m_{B}>0$ which in turn makes $m_{s}<0$. This is discussed in more details following Fig.~\ref{fig:mAmBnAnBlargeUlargeDelta}. For $\eta=1$ we see that there is a range of $U/\Delta$ for which there exists non-zero $m_{f}>0$. As we increase mass imbalance or equivalently decrease $\eta$, the spontaneously induced positive $m_{f}$ dies where as there emerges a mass imbalance induced $m_{f}<0$ in a wider parameter regime. Since, both non-zero $m_{s}$ and $m_{f}$ co-exist we can obtain a spin imbalanced ferromagnetic and/or ferrimagnetic phase in this regime. In (c) we show the single particle density of states at fermi level, $\rho(\omega \sim 0)$ for $\eta=0.4$. There exists two regimes of oppositely polarized AF half metals ($m_{f}=0$) separated in $U/\Delta$. The sliver of AF half-metal in low $U/\Delta$ regime is up spin polarized which is understandable because the heavier down spin species are immobile and located at the potential wells where as the lighter mass up spin species is available for conduction. As we increase correlation or equivalently decrease $\Delta$, the AF half metal becomes down spin polarized. In between the half metals, is a regime where both $\rho(\omega \sim 0)$ and $m_{f}$ are non-zero. This can be a spin imbalanced ferromagnetic and/or ferrimagnetic metallic phase which we will discuss in the following paragraph.   

\begin{figure}
    \centering
    
      \includegraphics[scale=0.245,angle=-90]{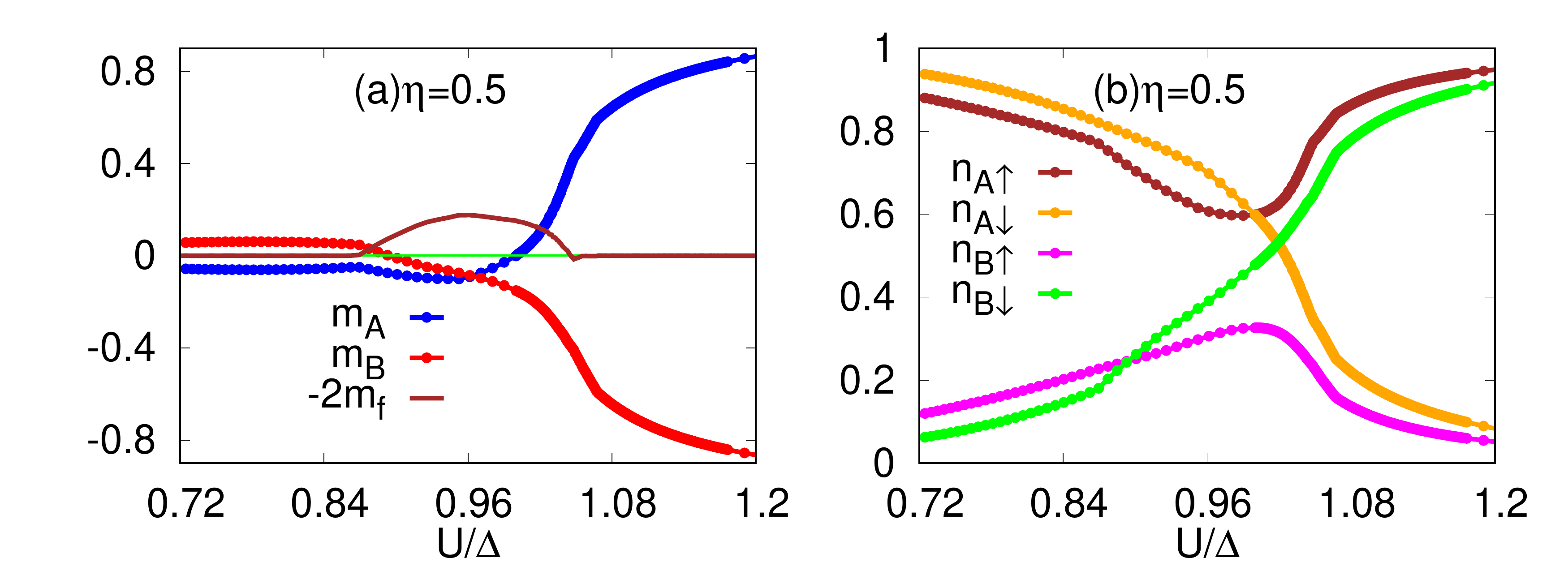}
      
  \includegraphics[scale=0.32]{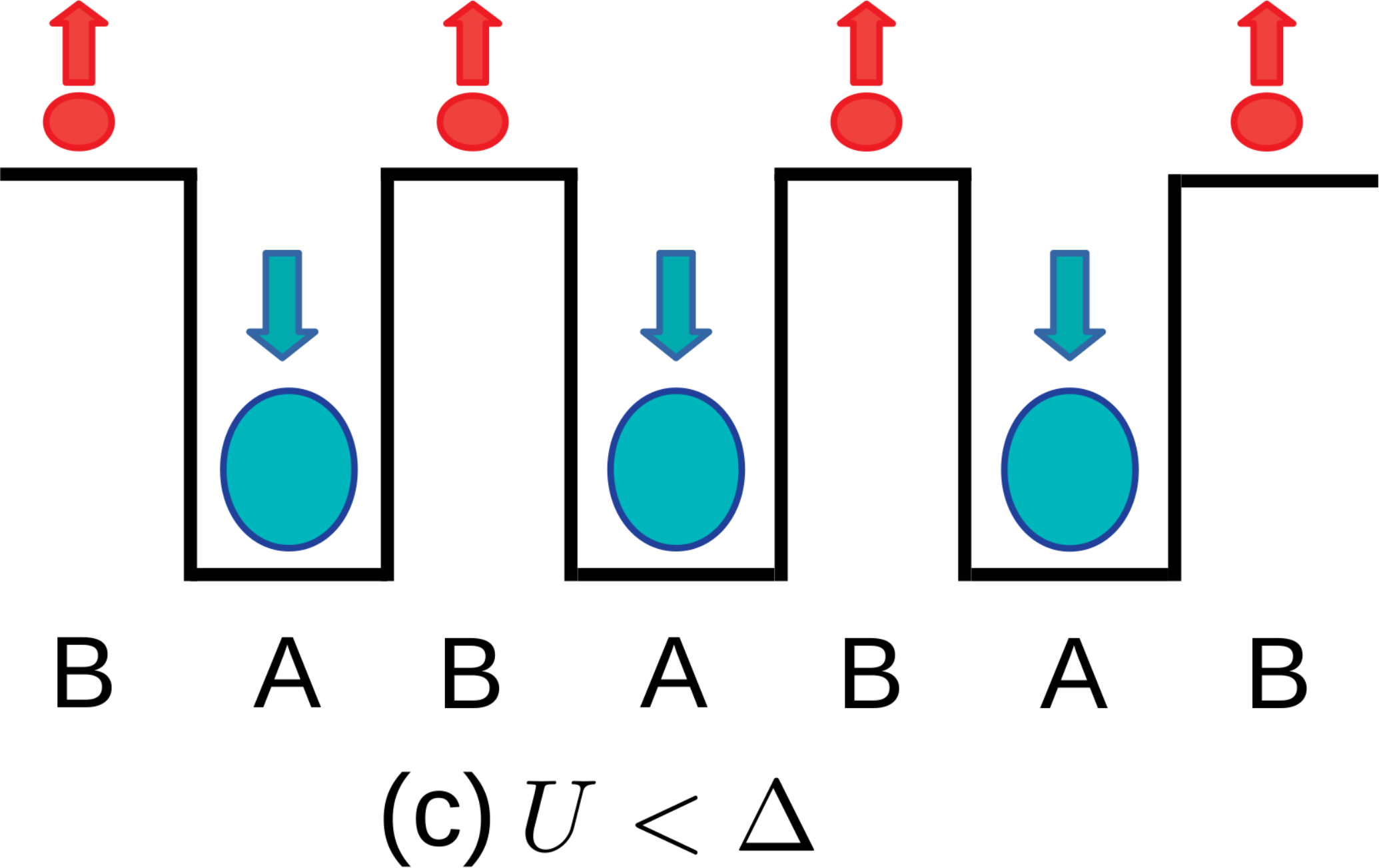}

    \caption{(a) shows the sublattice magnetization, $m_{\alpha},\alpha \in A,B$ for $U=10t_{\uparrow},\eta=0.5$. Also shown is negative of twice the uniform magnetization, $-2m_{f}$. For large $U/\Delta$, $m_{A}>0,m_{B}<0$. As we decrease $U/\Delta$, there is a regime where $m_{A}$ flips in sign but $m_{B}<0$. Here, $m_{A},m_{B}<0$. Finally, we have a regime where $m_{A}<0,m_{B}>0$. The regime of non-zero $m_{f}$ shows that between two ferrimagnetic phases, there exists a range of $U/\Delta$ where the system is spin imbalanced ferromagnetic. (b) shows the spin resolved densities on both sublattices for $\eta=0.5$ which in addition convey the existence of charge order throughout the parameter regime because of finite density difference between the sublattices. (c) gives a possible picture of magnetic ordering where $m_{A}<0,m_{B}>0$ for $U<\Delta$.} 
    \label{fig:mAmBnAnBlargeUlargeDelta}
\end{figure}

Fig.~\ref{fig:mAmBnAnBlargeUlargeDelta}(a) shows the sublattice magnetizations, $m_{A}$ and $m_{B}$ which show an interesting behaviour as we tune $U/\Delta$. Initially, for low values of $U/\Delta$, $m_{A}=-m_{B}$ with $m_{A}<0,m_{B}>0$ . In Fig.~\ref{fig:mAmBnAnBlargeUlargeDelta}(b), we see in this parameter regime, the down spin density is more than the up spin density on A sublattice which means the probability of occurrence of a down spin species is more than that of an up spin species on A sublattice. This is opposite for the B sublattice and can infact be visualized by a simple classical picture as shown in (c). Intuitively, the heavier mass down spin fermions prefer the potential wells as compared to the lighter mass up spin fermionic species. As we now increase $U/\Delta$, we reach a phase where $m_{f}\neq 0$ but $m_{A}<0,m_{B}>0$. This is a ferrimagnetic phase which we call, Ferri I phase. Further increasing $U/\Delta$, $m_{A}<0,m_{B}<0,$ and thereby $m_{f}\neq 0$ and we obtain a spin imbalanced ferromagnetic phase which is  purely ferromagnetic with $m_{A}=m_{B}$ at a point close to $U/\Delta \sim 1$. This leads to a second ferrimagnetic phase, which we call ferri II phase where $m_{A}>0,m_{B}<0$ and $m_{f} \neq 0$.  Ferri I,II and spin imbalanced ferro phases are all metallic in nature as seen from Fig.~\ref{fig:msmflargeUDelta}. Ultimately, upon further increasing $U/\Delta$, $m_{A}>0,m_{B}<0$ but $m_{f}=0$. The spin resolved densities in (b) assert that non zero density difference and magnetization co-exist throughout the parameter space. The  insulating phase with predominant spin ordering at large $U/\Delta$ is called a Mott insulating phase where as the insulating phase with predominant charge ordering at small $U/\Delta$ is called 
a correlated band insulator with weak AF SDW order.

\begin{figure}
    \centering
    \includegraphics[scale=0.245,angle=-90]{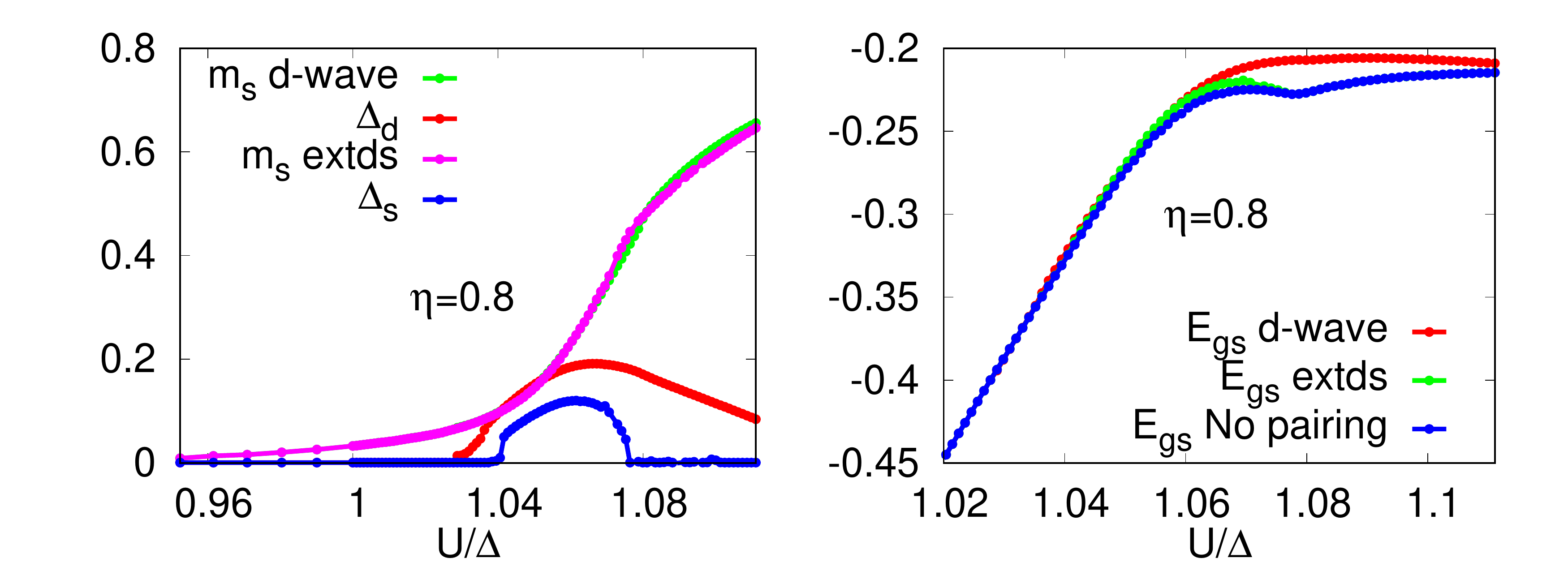}
    \caption{(a) shows the staggered magnetization, $m_{s}$ and pairing amplitude, $\Delta_{d,s}$ in the d-wave and extended s-wave pairing channels for $t_{\uparrow}'=0.45t_{\uparrow},\eta=0.8$. (b) shows the ground state energy comparison of the d-wave and extended s-wave symmetries with the state where pairing is not allowed. It shows that the d-wave and extended s-wave pairings which occur over a range of $U/\Delta$ are mostly meta-stable with the ground state energy of the no pairing state little lower than the pairing state.  }
    \label{fig:sc}
\end{figure}

\begin{figure}
    \centering
    \includegraphics[scale=0.32,angle=-90]{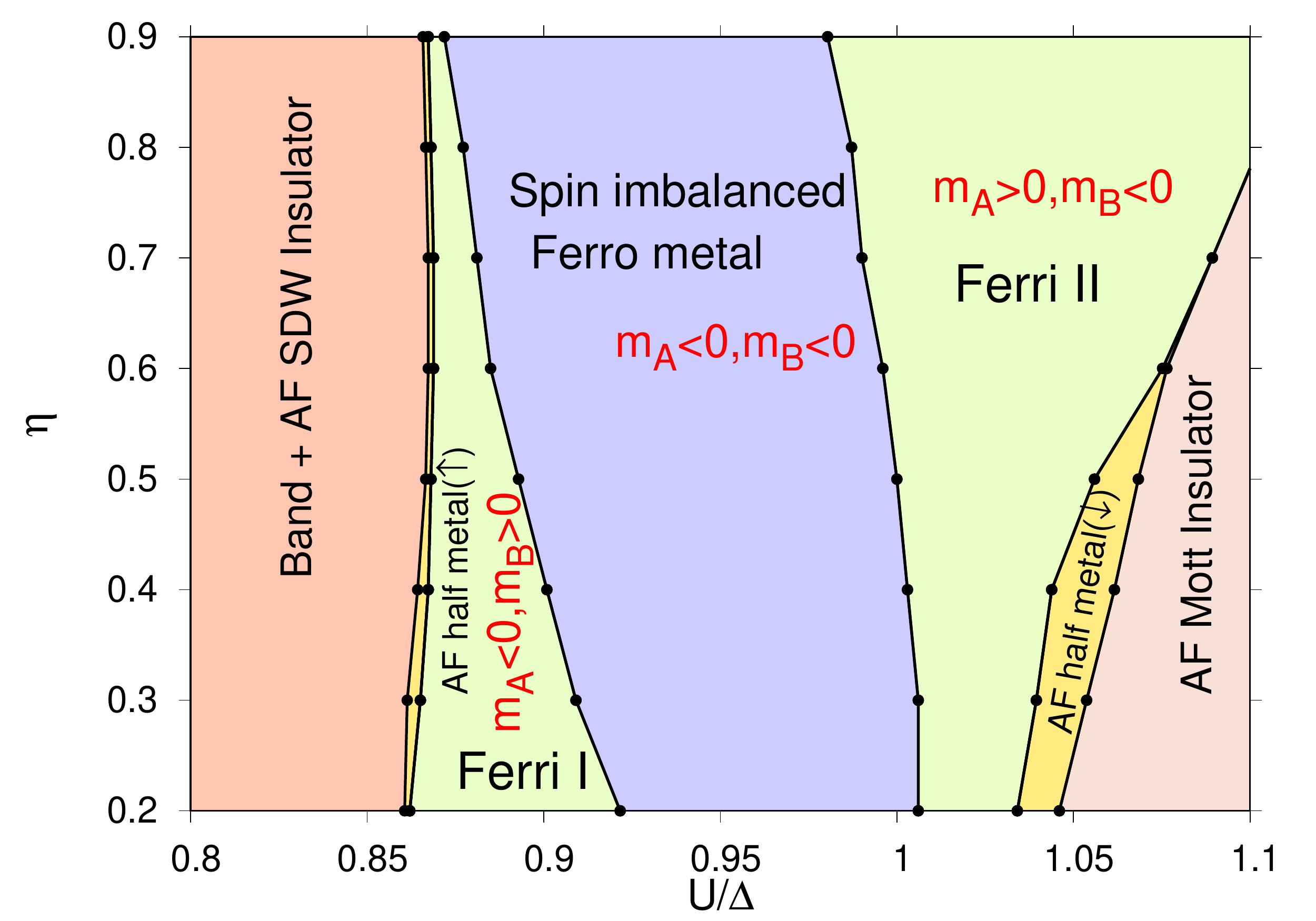}
    \caption{ Figure shows the phase diagram in $\eta-U/\Delta$ plane for $U=10t_{\uparrow},t_{\uparrow}'=0.3t_{\uparrow}$. Between two insulating phases namely, the AF Mott insulator and the correlated band insulator with weak AF SDW order, we have a broad range of exotic metallic phases bounded by AF half-metallic phases of opposite spin polarity. It consists of a Ferrimagnetic I phase where $m_{A}<0,m_{B}>0$, a novel spin imbalanced ferromagnetic phase with $m_{A},m_{B}<0$ and finally a Ferrimagnetic II phase where $m_{A}>0,m_{B}<0$.}
    \label{fig:pdlargeUlargeDelta}
\end{figure}

\begin{figure*}[htbp!]
    \centering
    \includegraphics[scale=0.35,angle=-90]{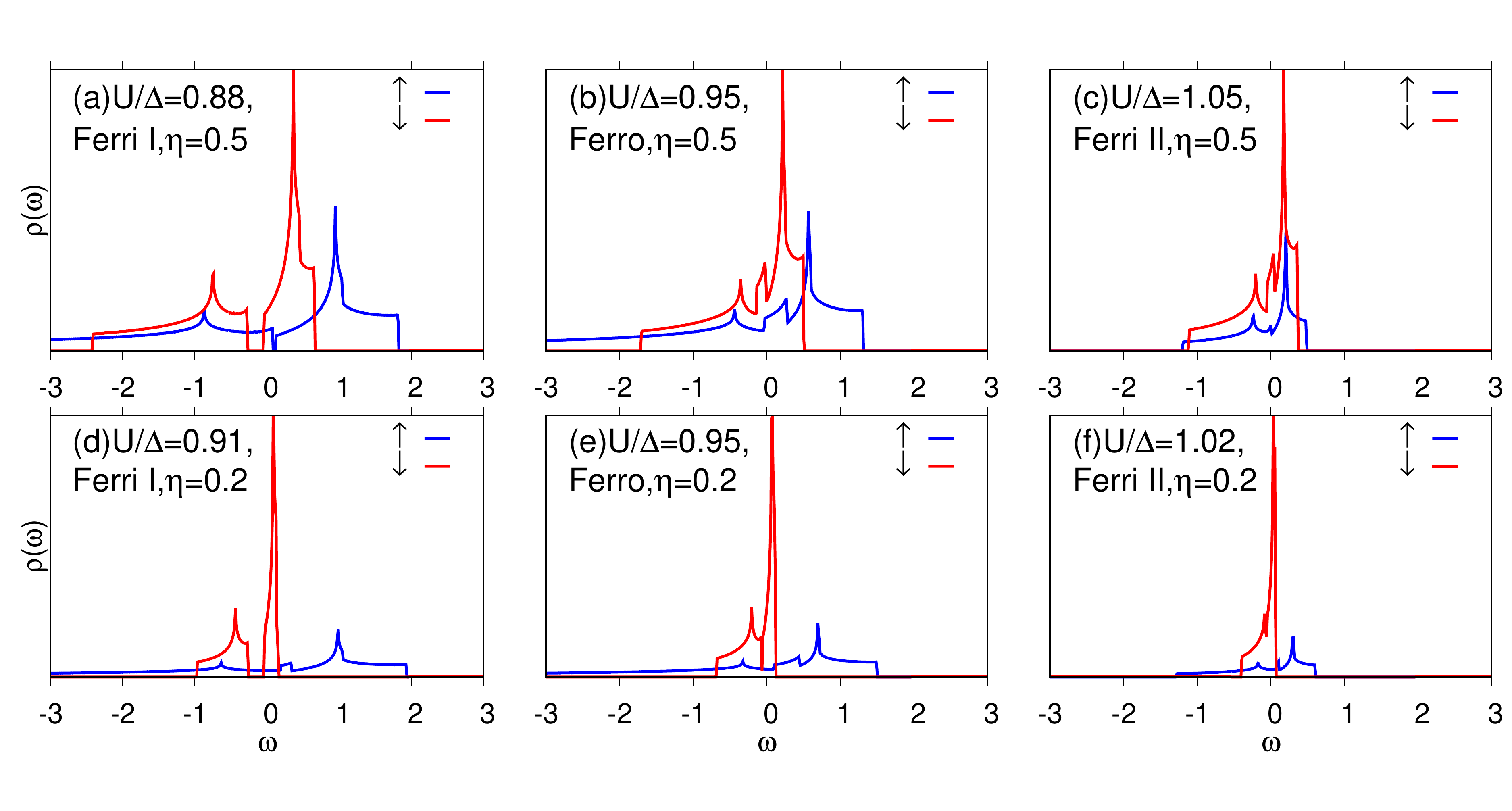}
    \caption{ Figure shows spin resolved single particle density of states, $\rho_{\sigma}(\omega)$ as function of $\omega$ for ferrimagnetic I, spin imbalanced ferromagnetic and ferrimagnetic II metallic phases for $\eta=0.5,0.2$. All phases have finite density of states at $\omega \sim 0$ in both spin channels which makes them metallic in nature. As mass imbalance is increased, effective band width in the down spin channel reduces and a large peak in the density of states in the down spin channel is observed near $\omega \sim 0$. }
    \label{fig:doslargeUlargeDelta}
\end{figure*}

We have investigated pairing in d-wave and extended s-wave channels in this limit by using a two step Bogoluibov de-Gennes calculation where we have solved for  staggered magnetization, $m_{s}$ and pairing amplitude, $\Delta_{d,s}$ simultaneously. In this method, it was assumned that inter band contribution to pairing is weak. For mass balanced case ~\cite{Anwesha3}, the model in this limit admits an unconventional superconducting phase sandwiched between exotic metallic phases. In Fig.~\ref{fig:sc}, we have shown the solution in both pairing channels for finite mass imbalance case. We found non-zero solution for both $m_{s}$ and  $\Delta_{d,s}$  for $\eta=0.8$. Here, $t_{\uparrow}'=0.45t_{\uparrow}$. The d-wave phase is wider than the extended s-wave case. However, the ground state energy comparison of the pairing states with the   
non-pairing state tells us that in most of the parameter space the superconductivity is metastable with ground state energy of the non-pairing state little lower than the pairing states. However, for low hopping asymmetries it can be expected that a regime of superconductivity coexisting with weak magnetic order is possible. Also due to the presence of large non-zero $m_{f}$, the Zeeman field is expected to be high which can give rise to exotic states like FFLO state where the SC order parameter is inhomogeneous in space. Other possibilities include breached pair or Sarma phase in the presence of mass imbalance.

Fig.~\ref{fig:pdlargeUlargeDelta} shows a rich phase diagram in $\eta-U/\Delta$ plane in the $U\sim\Delta \gg t_{\sigma},t_{\sigma}'$ limit. If we start from the low $U/\Delta$ regime, we are in the correlated band insulator with weak AF SDW order. As we increase $U/\Delta$, we first enter into a narrow sliver of AF half metal which is up spin polarized as the heavier down spin polarity fermions are immobile at the deep potential wells and the lighter up spin polarity fermionic species are available for conduction. On further increasing $U/\Delta$ we reach a ferrimagnetic metallic phase (Ferri I) where $m_{A}<0,m_{B}>0$. This phase is more dominant in the higher mass imbalance regime i.e., for lower values of $\eta$. Next, we enter a spin imbalanced ferromagnetic metallic phase where $m_{A},m_{B}<0$ which leads to a second ferrimagnetic metallic phase (Ferri II) for which $m_{A}>0,m_{B}<0$. Ferri II phase is more dominant for higher values of $\eta$. Ultimately, on further increasing $U/\Delta$, we enter into the Mott insulating phase through a phase of down spin polarized AF half-metal. Thus, the system does not only show charge dynamics but also novel spin dynamics.

\begin{figure}[htbp!]
    \centering
    \includegraphics[scale=0.33,angle=-90]{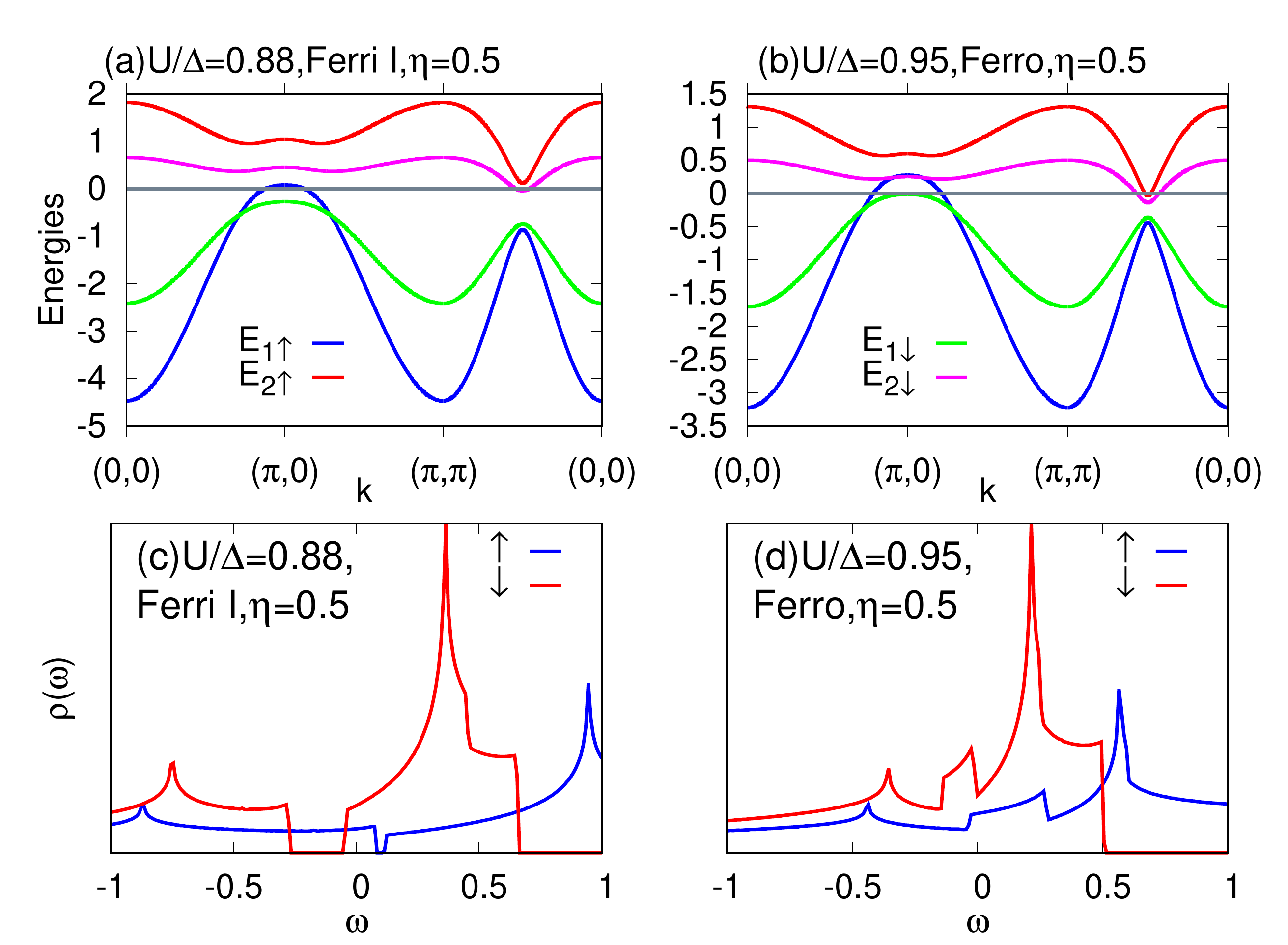}
    \caption{(a) and (b) show the band dispersions along the high symmetry directions in the Brillouin zone for the Ferri I and Ferro phases respectively. (c) and (d) show the density of states for the same two phases, zoomed around $\omega \sim 0$. The fermi level in the ferrimagnetic phase lies within the up spin polarity valence band and the down spin polarity conduction band consistent with the band picture where $E_{1\uparrow}$ crosses the fermi level to become partially unoccupied (hole pockects) around $(\pi,0)$ and $E_{2\downarrow}$ crosses the fermi level to get partially occupied resulting in electron pockets around $(\pi/2,\pi/2)$. For the ferromagnetic phase, fermi level lies in the region where both valence and conduction band merge for the up spin channel where as in the down spin channel, the conduction band only crosses the fermi level, with the valence band just a little below the fermi level.   }
    \label{fig:displargeUlargeDelta}
\end{figure}

\begin{figure*}[htbp!]
    \centering
    \includegraphics[scale=0.35,angle=-90]{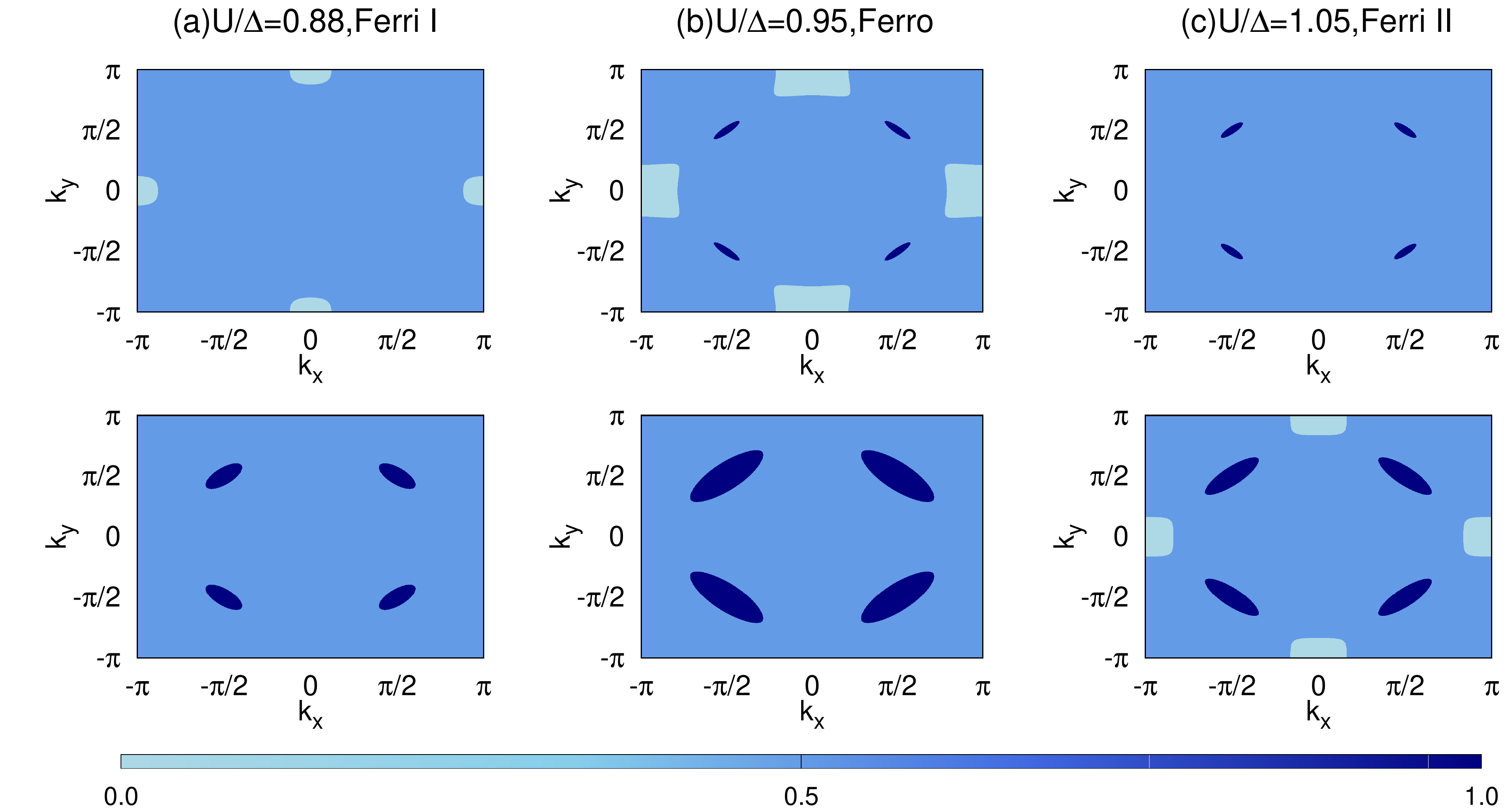}
    \caption{ Figure shows the momentum distribution function, $n_{k\sigma}$ in the full Brillouin zone for the ferrimagnetic I phase, spin imbalanced ferromagnetic phase and the ferrimagnetic II phase for $\eta=0.5$. The upper row represents the up spin channel and lower row the down spin channel. The ferri I phase has hole pockets ($n_{k\uparrow}<1/2$) in the up spin channel and electron pockets ($n_{k\downarrow}>1/2$) in the down spin channel. The ferro phase has large hole pockets and tiny electron pockets in the up spin channel and only large electron pockets in the down spin channel. The ferri II phase on the other hand has only electron pockets in the up spin channel and both electron and hole pockets in the down spin channel. The electron pockets occur around $(\pm \pi/2,\pm \pi/2)$ and hole pockets occur around $(0,\pm \pi),(\pm \pi,0)$.}
    \label{fig:nklargeUlargeDelta}
\end{figure*}

Fig.~\ref{fig:doslargeUlargeDelta} shows the single particle density of states, $\rho_{\sigma}(\omega)=1/2\sum_{\alpha}\rho_{\alpha\sigma},\alpha \in A,B$ where $\rho_{\alpha\sigma}=-\sum_{k} \text{Im} G_{\alpha\sigma}(k,\omega^{+})/\pi$. The single particle Green's function, $G_{\alpha\sigma}(k,\omega)$ is renormalized by suitable Gutzwiller factors. The bandwidth of both spin polarities changes as we tune $U/\Delta$ due to the renormalization of the kinetic energy by Gutzwiller factors which are themselves functions of $U/\Delta$. In all three phases viz, the Ferrimagnetic I, spin imbalanced ferromagnetic and the Ferrimagnetic II phases, there exists finite $\rho(\omega \sim 0)$  making them metallic in nature. Due to the mass imbalance the effective bandwidth of the down spin channel shrinks in comparison to the up spin channel. As mass imbalance grows, the down spin polarity bands become thinner and sharper giving a quasi flat band structure to the bands. This is crucial for the stabilization of ferromagnetism along with frustration as pointed out earlier. The spin imbalanced ferromagnetic phase is unique to the mass imbalanced case and does not exist in the mass balanced version unlike the ferrimagnetic metallic phase which exists even for $\eta=1$.

Fig.~\ref{fig:displargeUlargeDelta} shows the band dispersions for the ferrimagnetic I and ferromagnetic phases and also the density of states which are zoomed near $\omega \sim 0$. There are in total four bands : two corresponding to two different sublattices which we are calling valence and conduction bands which again come in spin up and spin down variants. In the ferrimagnetic phase the up spin polarity valence band, $E_{1\uparrow}$ crosses the fermi level, thus getting partially unoccupied. It results into hole pockects around $(0,\pm \pi),(\pm \pi,0)$. This can further be seen from (c) where the fermi level is found to be lying inside the valence band for up spin polarity. Also, the conduction band of down spin polarity, $E_{2\downarrow}$ crosses the fermi level to become partially occupied creating electron pockets around $(\pm \pi/2,\pm \pi/2)$. This is also seen in density of states plot where the fermi level lies inside the down spin polarity conduction band. In case of the spin imbalanced ferromagnetic phase, the fermi level lies in the regime where both valence and conduction band of the up spin channel overlap and contribute to the density of states. Correspondingly both $E_{1\uparrow}$ and $E_{2\uparrow}$ cross the fermi level creating hole and electron pockets respectively. In the down spin channel, the valence and the conduction band overlap and contribute to the density of states at the same $\omega$ value but it occurs for a range of $\omega$ values just below the fermi level. The fermi level in this case is inside the conduction band and has corresponding electron pockets. The down spin polarity valence band, $E_{1\downarrow}$ is very near to the fermi level but does not cross it. 

Fig.~\ref{fig:nklargeUlargeDelta} shows the momentum distribution function, $n_{k\sigma}=1/2\sum_{\alpha}\int_{-\infty}^{0}A_{\alpha\sigma}(k,\omega)d\omega$ where, $A_{\alpha\sigma}(k,\omega)$ is the spin resolved single particle spectral function for $\alpha \in A,B$ sublattice. Here, $\eta=0.5$. In this figure, electron pockets ($n_{k\sigma}>1/2$) occur around $(\pm \pi/2,\pm \pi/2)$ and hole pockets around $(0,\pm \pi),(\pm \pi,0)$. In the ferri I phase, the up spin channel has only hole pockets and down spin channel has only electron pockets. This is also ascertained by Fig.~\ref{fig:displargeUlargeDelta}. This means the density in the up channel which is nothing but the sum of  $n_{k\uparrow}$ over the Brillouin zone is lesser than the density in the down spin channel, which makes $m_{f}=n_{\uparrow}-n_{\downarrow}<0$. This we have already seen earlier for the ferri I phase. Similarly for the ferro phase we see, that the up channel has large hole pockets and tiny electron pockets and the down spin channel on the other hand has large electron pockets. This again makes $m_{f}<0$ which was also seen earlier for the ferro phase. In the ferri II phase , up spin channel has electron pockets but no hole pockets where as the down spin channel has both electron and hole pockets. The hole pockets in the down spin channel has to be little larger than the electron pockets so that density in the down spin channel is $<1/2$ so as to maintain half-filling constraint since the up spin channel has only electron pockets making density in the up spin channel $>1/2$. This makes $m_{f}>0$ and indeed there is a narrow regime for $\eta=0.5$ where $m_{A}>0,m_{B}<0,m_{A}>|m_{B}|$. But as can be understood $m_{f}$ is small in magnitude here.

\section{Conclusion}
We have presented the results for the mass imbalanced frustrated IHM in the low to intermediate range of $U,\Delta$ using unrestricted Hartree-Fock theory and also in the $U\sim \Delta \gg t_{\sigma},t_{\sigma}'$ limit using generalized Gutzwiller renormalized mean field theory. We have also qualitatively discussed about the limit $U\gg \Delta,t_{\sigma},t_{\sigma}'$. The former two limits admit novel magnetic metallic phases and has potential applications in the field of spintronics. In addition, in the strong coupling limit, a metastable singlet superconducting phase was observed in the presence of mass imbalance. However, among all these phases, the spin imbalanced ferromagnetic metallic phase is an outcome of subtle interplay between mass imbalance and frustration. Moreover, there is significant broadening of the ferrimagnetic phase in the mass imbalanced case as compared to the mass balanced case. Also in the strong coupling limit, AF half metals of opposite spin polarities occur separated in the phase diagram by $U/\Delta$, which means we can switch from up spin polarized conduction to down spin polarized conduction and vice-versa just by tuning $U/\Delta$ appropriately. As already pointed out that the variation of $\eta$ can be done by lattice depth tuning in an optical lattice and thus mass imbalanced fermions on an optical lattice can be used to scan our entire phase diagram in both limits. Further, layered materials like graphene on h-BN substrate which can show staggered ionicity due to the difference in the energies of Boron and Nitrogen and which may possess bands with different bandwidths with possibility of nearly flat bands is expected to show physics of this model. 

{\it Acknowledgements}. The author would like to thank Arti Garg for discussions. The author was funded by Department of Atomic Energy, India.

\section*{Appendix A}

Here, we discuss the details of the unrestricted Hartree-Fock theory used in the low to intermediate regimes of $U,\Delta$. We decompose the Hubbard $U$ term in terms of spin resolved densities, $n_{i\alpha\sigma}$ where $\alpha \in A,B$ sublattices.

\begin{align}
Un_{i\alpha\uparrow}n_{i\alpha\downarrow}\approx U\langle n_{i\alpha\uparrow} \rangle n_{i\alpha\downarrow}+Un_{i\alpha\uparrow}\langle n_{i\alpha\downarrow}\rangle-U\langle n_{i\alpha\uparrow}\rangle \langle n_{i\alpha\downarrow} \rangle   
\end{align}

The effective mean field Hamiltonian,$H_{eff}$ is given by,

\begin{align}
    &H_{eff}=\sum_{k\sigma}\bigg(\dfrac{U(1+\delta-\sigma m_{A})}{2}-\dfrac{\Delta}{2}-t_{\sigma}'\gamma_{k}^{'}-\mu\bigg)c_{kA\sigma}^{\dagger}c_{kA\sigma}\nonumber\\&+\bigg(\dfrac{U(1-\delta-\sigma m_{B})}{2}+\dfrac{\Delta}{2}-t_{\sigma}'\gamma_{k}^{'}-\mu\bigg)c_{kB\sigma}^{\dagger}c_{kB\sigma}\nonumber\\&+t_{\sigma}\gamma_{k}(c_{kA\sigma}^{\dagger}c_{kB\sigma} +h.c.)
\end{align}

where, $\gamma_{k}=2(\cos(k_{x})+\cos(k_{y}))$ and $\gamma_{k}'=4\cos(k_{x})\cos(k_{y})$.

We diagonalize the Hamiltonian by using the following canonical transformation,
\begin{equation}
    \begin{aligned}
    &c_{kA\sigma}=\alpha_{k\sigma}d_{k1\sigma}+\beta_{k\sigma}d_{k2\sigma}\\ &c_{kB\sigma}=\alpha_{k\sigma}d_{k2\sigma}-\beta_{k\sigma}d_{k1\sigma}
\end{aligned}
\end{equation}

where, $\alpha_{k\sigma}^{2}=\frac{1}{2}[1-\frac{\tilde{\Delta}_{\sigma}}{\sqrt{\tilde{\Delta}_{\sigma}^{2}+t_{\sigma}^{2}\gamma_{k}^{2}}}]$ and $\beta_{k\sigma}^{2}=\frac{1}{2}[1+\frac{\tilde{\Delta}_{\sigma}}{\sqrt{\tilde{\Delta}_{\sigma}^{2}+t_{\sigma}^{2}\gamma_{k}^{2}}}]$. Here, the effective ionic potential, $\tilde{\Delta}_{\sigma}$ felt by the system is given by $\tilde{\Delta}_{\sigma}=\dfrac{U(\delta-\sigma m_{s})-\Delta}{2}$.

The eigen energies, $\lambda^{1,2}_{\sigma}$, are given by,

\begin{align}
   \lambda^{1,2}_{\sigma}=\bigg[\dfrac{U(1-\sigma m_{f})}{2}-\mu-t_{\sigma}'\gamma_{k}' \bigg]\mp \sqrt{\tilde{\Delta}_{\sigma}^{2}+t_{\sigma}^{2}\gamma_{k}^{2}} 
\end{align}

The self-consistent equations of the spin resolved densities, $n_{\alpha\sigma}$ is given by,
\begin{equation}
    \begin{aligned}
    n_{A\sigma}=\dfrac{1}{N^{2}}\sum_{k\in FBZ}[\alpha_{k\sigma}^{2}\langle d_{k1\sigma}^{\dagger}d_{k1\sigma}\rangle+\beta_{k\sigma}^{2}\langle d_{k2\sigma}^{\dagger}d_{k2\sigma}\rangle]\\
    n_{B\sigma}=\dfrac{1}{N^{2}}\sum_{k\in FBZ}[\alpha_{k\sigma}^{2}\langle d_{k2\sigma}^{\dagger}d_{k2\sigma}\rangle+\beta_{k\sigma}^{2}\langle d_{k1\sigma}^{\dagger}d_{k1\sigma}\rangle]
\end{aligned}
\end{equation}

Here the sum is over full Brillouin zone (FBZ). The occupation probability of the bands is given by the Fermi Dirac distribution. For zero temperature, this implies that the bands will be occupied if their energies, $\lambda^{1,2}_{\sigma}$ are below the fermi level. From the spin resolved densities, we can construct linear combinations like sublattice magnetization, $m_{\alpha}=n_{\alpha\uparrow}-n_{\alpha\downarrow}$ and density difference between sublattices, $\delta=(n_{A}-n_{B})/2$, which are of physical interest.

\section*{Appendix B}

In the limit $U\sim \Delta \gg t_{\sigma},t_{\sigma}'$, we construct the effective low energy Hamiltonian by doing a site dependent projection of holes from A sublattice having ionic potential -$\Delta/2$ and doublons from B sublattice having ionic potential $\Delta/2$. There is a term in the effective Hamiltonian, as shown in main text, which is purely induced by mass imbalance and is $\propto (t_{\uparrow}^{2}-t_{\downarrow}^{2})$. Here, we show the derivation of this term starting from the Hubbard operators, $X_{i\alpha}^{\psi \leftarrow \phi}$ which creates state $\psi$ from state $\phi$ on the $i-th$ site belonging to $\alpha \in A,B$ sublattice. States $\phi,\psi \in |0\rangle,|\uparrow\rangle,|\downarrow\rangle,|\uparrow\downarrow\rangle$.

The commutator in the effective Hamiltonian $\frac{1}{U+\Delta}[{H_{t}^{+}}_{A\rightarrow B},{H_{t}^{-}}_{B\rightarrow A}]$ represents two site processes which either flip two oppositely oriented spins on neighbouring A and B sites or preserve them through a virtual high energy state consisting of hole on A site and doublon on B site. The spin flip process correspond to the $(S_{iA}^{x}S_{jB}^{x}+S_{iA}^{y}S_{jB}^{y})$ term where as the spin preserving process corresponds to two terms : the $S_{iA}^{z}S_{jB}^{z}$ term and the mass imbalance induced staggered magnetic field term. The relevant spin preserving term can be written as,
$-\frac{t_{\sigma}^{2}}{U+\Delta}X_{iA}^{\sigma \leftarrow \sigma}X_{jB}^{\bar{\sigma} \leftarrow \bar{\sigma}}$
where, $X_{iA}^{\sigma \leftarrow \sigma}=\frac{2-n_{iA}}{2}+\sigma S_{iA}^{z}$ in the space where holes are not allowed and $X_{jB}^{\sigma \leftarrow \sigma}=\dfrac{n_{jB}}{2}+\sigma S_{jB}^{z}$
in the space where doublons are not allowed.

Putting back these expressions, we get $\frac{t_{\uparrow}^{2}+t_{\downarrow}^{2}}{U+\Delta}[S_{iA}^{z}S_{jB}^{z}-\frac{(2-n_{iA})n_{jB}}{4}]+\frac{t_{\uparrow}^{2}-t_{\downarrow}^{2}}{U+\Delta}[\frac{2-n_{iA}}{2}S_{jB}^{z}-\frac{n_{jB}}{2}S_{iA}^{z}]$. The latter term is the mass imbalance induced staggered magnetic field term which becomes zero in the mass balanced case.

After we obtain the low energy effective Hamiltonian, we renormalize the couplings with suitable weight factors known as Gutzwiller factors which take care of the site dependent projection approximately. The Gutzwiller factors used in the calculation are listed in Table ~\ref{table:gutz}.

\large
\begin{table}[h!]
\renewcommand{\arraystretch}{2.2}
 \begin{tabular}{||c | c ||} 
 \hline
 GF & Expressions\\[2ex]
 \hline
$ g_{t\sigma}$ &  $\dfrac{2\delta}{\sqrt{(1+\delta+\sigma m_{A})(1+\delta-\sigma m_{B})}}$\\[1.5ex] 
\hline
$g_{A\sigma}$ & $\dfrac{2\delta}{1+\delta+\sigma m_{A}}$\\ [1.5ex] \hline
$g_{B\sigma}$ & $\dfrac{2\delta}{1+\delta-\sigma m_{B}}$\\[1.5ex]  \hline
$g_{s\alpha_{1}\alpha_{2}}$  &  $\dfrac{4}{\sqrt{((1+\delta)^2-m_{\alpha_{1}}^2)((1+\delta)^2-m_{\alpha_{2}}^2)}}$ \\[1.5ex] \hline
$g_{1}$  & $\delta g_{sAB}$ \\ \hline
$g_{\alpha_{1}\alpha_{1}\alpha_{2}\sigma}$ & $\dfrac{4\delta}{\sqrt{((1+\delta)^{2}-m_{\alpha_{1}}^{2})(1+\delta+\sigma m_{\alpha_{1}})(1+\delta+\sigma m_{\alpha_{2}})}}$\\[-1.5ex]
$(\alpha_{1} \neq \alpha_{2})$ &\\[1ex]
\hline
\end{tabular}
\renewcommand{\arraystretch}{0.5}

\caption{Expressions of Gutzwiller factors at half-filling where $\alpha_{1,2}\in A,B$.}
\label{table:gutz}
\end{table}

\normalsize

$g_{t\sigma}$ renormalizes the nearest neighbor hopping and the $t_{\sigma}t_{\sigma}'$ effective hopping between A and B sublattice. $g_{A\sigma}$ renormalizes the effective hopping of doublons on A sublattice and $g_{B\sigma}$ renormalizes the effective hopping of holes on B sublattice. Both these processes preserve the polarity of spin at the intermediate site via which these three site hopping processes are accomplished. $g_{s\alpha_{1}\alpha_{2}}$, where $\alpha_{1},\alpha_{2}\in A,B$, renormalizes the spin flip part of the inter or intra sublattice Heisenberg term. The dimer terms involving product of density operators and the spin preserving part of the Heisenberg term are renormalized by unity. The trimer terms which involve a flip in spin polarity on the intermediate site accompanied by effective intra sublattice hopping of doublons or holes are renormalized by $g_{1}$. $g_{AAB\bar{\sigma}},g_{BBA\sigma}$ renormalize  the corresponding $t_{\sigma}t_{\bar{\sigma}}'$ processes.

\end{document}